\documentclass[%
reprint,
superscriptaddress,
nofootinbib,
amsmath,amssymb,
aps,
]{revtex4-1}

\usepackage{graphicx}
\usepackage{dcolumn}
\usepackage{bm}
\usepackage{xcolor}

\usepackage{subfigure}

\begin{document}
	
\preprint{APS/123-QED}

\title{Search for gravitational waves from Scorpius X-1 in the first Advanced LIGO observing run with a hidden Markov model}

\author{%
	B.~P.~Abbott,$^{1}$  
	R.~Abbott,$^{1}$  
	T.~D.~Abbott,$^{2}$  
	F.~Acernese,$^{3,4}$ 
	K.~Ackley,$^{5}$  
	C.~Adams,$^{6}$  
	T.~Adams,$^{7}$ 
	P.~Addesso,$^{8}$  
	R.~X.~Adhikari,$^{1}$  
	V.~B.~Adya,$^{9}$
% ...
	\\
	\medskip
	(LIGO Scientific Collaboration and Virgo Collaboration) 
	\\
	\medskip
	{{}$^{*}$Deceased, March 2016. }%
	{{}$^{\sharp}$Deceased, March 2017. }%
	{${}^{\dag}$Deceased, February 2017. }%
	{${}^{\ddag}$Deceased, December 2016. }%
	\\
	\medskip
	S.~Suvorova,$^{132,148}$  %
	W.~Moran,$^{148}$  %
	and
	R.~J.~Evans$^{132}$
}\noaffiliation
\affiliation {LIGO, California Institute of Technology, Pasadena, CA 91125, USA }
\affiliation {Louisiana State University, Baton Rouge, LA 70803, USA }
\affiliation {Universit\`a di Salerno, Fisciano, I-84084 Salerno, Italy }
\affiliation {INFN, Sezione di Napoli, Complesso Universitario di Monte S.Angelo, I-80126 Napoli, Italy }
\affiliation {University of Florida, Gainesville, FL 32611, USA }
\affiliation {LIGO Livingston Observatory, Livingston, LA 70754, USA }
\affiliation {Laboratoire d'Annecy-le-Vieux de Physique des Particules (LAPP), Universit\'e Savoie Mont Blanc, CNRS/IN2P3, F-74941 Annecy, France }
\affiliation {University of Sannio at Benevento, I-82100 Benevento, Italy and INFN, Sezione di Napoli, I-80100 Napoli, Italy }
\affiliation {Albert-Einstein-Institut, Max-Planck-Institut f\"ur Gravi\-ta\-tions\-physik, D-30167 Hannover, Germany }
\affiliation {The University of Mississippi, University, MS 38677, USA }
\affiliation {NCSA, University of Illinois at Urbana-Champaign, Urbana, IL 61801, USA }
\affiliation {Nikhef, Science Park, 1098 XG Amsterdam, The Netherlands }
\affiliation {LIGO, Massachusetts Institute of Technology, Cambridge, MA 02139, USA }
\affiliation {Instituto Nacional de Pesquisas Espaciais, 12227-010 S\~{a}o Jos\'{e} dos Campos, S\~{a}o Paulo, Brazil }
\affiliation {Gran Sasso Science Institute (GSSI), I-67100 L'Aquila, Italy }
\affiliation {INFN, Sezione di Roma Tor Vergata, I-00133 Roma, Italy }
\affiliation {Inter-University Centre for Astronomy and Astrophysics, Pune 411007, India }
\affiliation {International Centre for Theoretical Sciences, Tata Institute of Fundamental Research, Bengaluru 560089, India }
\affiliation {University of Wisconsin-Milwaukee, Milwaukee, WI 53201, USA }
\affiliation {Leibniz Universit\"at Hannover, D-30167 Hannover, Germany }
\affiliation {Universit\`a di Pisa, I-56127 Pisa, Italy }
\affiliation {INFN, Sezione di Pisa, I-56127 Pisa, Italy }
\affiliation {SUPA, University of Glasgow, Glasgow G12 8QQ, United Kingdom }
\affiliation {Australian National University, Canberra, Australian Capital Territory 0200, Australia }
\affiliation {Laboratoire des Mat\'eriaux Avanc\'es (LMA), CNRS/IN2P3, F-69622 Villeurbanne, France }
\affiliation {LAL, Univ. Paris-Sud, CNRS/IN2P3, Universit\'e Paris-Saclay, F-91898 Orsay, France }
\affiliation {California State University Fullerton, Fullerton, CA 92831, USA }
\affiliation {European Gravitational Observatory (EGO), I-56021 Cascina, Pisa, Italy }
\affiliation {Chennai Mathematical Institute, Chennai 603103, India }
\affiliation {Universit\`a di Roma Tor Vergata, I-00133 Roma, Italy }
\affiliation {Universit\"at Hamburg, D-22761 Hamburg, Germany }
\affiliation {INFN, Sezione di Roma, I-00185 Roma, Italy }
\affiliation {Embry-Riddle Aeronautical University, Prescott, AZ 86301, USA }
\affiliation {Albert-Einstein-Institut, Max-Planck-Institut f\"ur Gravitations\-physik, D-14476 Potsdam-Golm, Germany }
\affiliation {APC, AstroParticule et Cosmologie, Universit\'e Paris Diderot, CNRS/IN2P3, CEA/Irfu, Observatoire de Paris, Sorbonne Paris Cit\'e, F-75205 Paris Cedex 13, France }
\affiliation {Korea Institute of Science and Technology Information, Daejeon 34141, Korea }
\affiliation {West Virginia University, Morgantown, WV 26506, USA }
\affiliation {Center for Gravitational Waves and Cosmology, West Virginia University, Morgantown, WV 26505, USA }
\affiliation {Universit\`a di Perugia, I-06123 Perugia, Italy }
\affiliation {INFN, Sezione di Perugia, I-06123 Perugia, Italy }
\affiliation {Syracuse University, Syracuse, NY 13244, USA }
\affiliation {University of Minnesota, Minneapolis, MN 55455, USA }
\affiliation {LIGO Hanford Observatory, Richland, WA 99352, USA }
\affiliation {Wigner RCP, RMKI, H-1121 Budapest, Konkoly Thege Mikl\'os \'ut 29-33, Hungary }
\affiliation {Columbia University, New York, NY 10027, USA }
\affiliation {Stanford University, Stanford, CA 94305, USA }
\affiliation {Universit\`a di Camerino, Dipartimento di Fisica, I-62032 Camerino, Italy }
\affiliation {Universit\`a di Padova, Dipartimento di Fisica e Astronomia, I-35131 Padova, Italy }
\affiliation {INFN, Sezione di Padova, I-35131 Padova, Italy }
\affiliation {MTA E\"otv\"os University, ``Lendulet'' Astrophysics Research Group, Budapest 1117, Hungary }
\affiliation {Nicolaus Copernicus Astronomical Center, Polish Academy of Sciences, 00-716, Warsaw, Poland }
\affiliation {University of Birmingham, Birmingham B15 2TT, United Kingdom }
\affiliation {Universit\`a degli Studi di Genova, I-16146 Genova, Italy }
\affiliation {INFN, Sezione di Genova, I-16146 Genova, Italy }
\affiliation {RRCAT, Indore MP 452013, India }
\affiliation {Faculty of Physics, Lomonosov Moscow State University, Moscow 119991, Russia }
\affiliation {SUPA, University of the West of Scotland, Paisley PA1 2BE, United Kingdom }
\affiliation {Caltech CaRT, Pasadena, CA 91125, USA }
\affiliation {University of Western Australia, Crawley, Western Australia 6009, Australia }
\affiliation {Department of Astrophysics/IMAPP, Radboud University Nijmegen, P.O. Box 9010, 6500 GL Nijmegen, The Netherlands }
\affiliation {Artemis, Universit\'e C\^ote d'Azur, Observatoire C\^ote d'Azur, CNRS, CS 34229, F-06304 Nice Cedex 4, France }
\affiliation {Institut de Physique de Rennes, CNRS, Universit\'e de Rennes 1, F-35042 Rennes, France }
\affiliation {Washington State University, Pullman, WA 99164, USA }
\affiliation {Universit\`a degli Studi di Urbino 'Carlo Bo', I-61029 Urbino, Italy }
\affiliation {INFN, Sezione di Firenze, I-50019 Sesto Fiorentino, Firenze, Italy }
\affiliation {University of Oregon, Eugene, OR 97403, USA }
\affiliation {Laboratoire Kastler Brossel, UPMC-Sorbonne Universit\'es, CNRS, ENS-PSL Research University, Coll\`ege de France, F-75005 Paris, France }
\affiliation {Carleton College, Northfield, MN 55057, USA }
\affiliation {Astronomical Observatory Warsaw University, 00-478 Warsaw, Poland }
\affiliation {VU University Amsterdam, 1081 HV Amsterdam, The Netherlands }
\affiliation {University of Maryland, College Park, MD 20742, USA }
\affiliation {Center for Relativistic Astrophysics and School of Physics, Georgia Institute of Technology, Atlanta, GA 30332, USA }
\affiliation {Universit\'e Claude Bernard Lyon 1, F-69622 Villeurbanne, France }
\affiliation {Universit\`a di Napoli 'Federico II', Complesso Universitario di Monte S.Angelo, I-80126 Napoli, Italy }
\affiliation {NASA Goddard Space Flight Center, Greenbelt, MD 20771, USA }
\affiliation {RESCEU, University of Tokyo, Tokyo, 113-0033, Japan. }
\affiliation {University of Adelaide, Adelaide, South Australia 5005, Australia }
\affiliation {Tsinghua University, Beijing 100084, China }
\affiliation {Texas Tech University, Lubbock, TX 79409, USA }
\affiliation {Kenyon College, Gambier, OH 43022, USA }
\affiliation {The Pennsylvania State University, University Park, PA 16802, USA }
\affiliation {National Tsing Hua University, Hsinchu City, 30013 Taiwan, Republic of China }
\affiliation {Charles Sturt University, Wagga Wagga, New South Wales 2678, Australia }
\affiliation {University of Chicago, Chicago, IL 60637, USA }
\affiliation {Pusan National University, Busan 46241, Korea }
\affiliation {University of Cambridge, Cambridge CB2 1TN, United Kingdom }
\affiliation {The Chinese University of Hong Kong, Shatin, NT, Hong Kong }
\affiliation {INAF, Osservatorio Astronomico di Padova, Vicolo dell'Osservatorio 5, I-35122 Padova, Italy }
\affiliation {INFN, Trento Institute for Fundamental Physics and Applications, I-38123 Povo, Trento, Italy }
\affiliation {Universit\`a di Roma 'La Sapienza', I-00185 Roma, Italy }
\affiliation {Universit\'e Libre de Bruxelles, Brussels 1050, Belgium }
\affiliation {Sonoma State University, Rohnert Park, CA 94928, USA }
\affiliation {Montana State University, Bozeman, MT 59717, USA }
\affiliation {Center for Interdisciplinary Exploration \& Research in Astrophysics (CIERA), Northwestern University, Evanston, IL 60208, USA }
\affiliation {Cardiff University, Cardiff CF24 3AA, United Kingdom }
\affiliation {Universitat de les Illes Balears, IAC3---IEEC, E-07122 Palma de Mallorca, Spain }
\affiliation {The University of Texas Rio Grande Valley, Brownsville, TX 78520, USA }
\affiliation {Bellevue College, Bellevue, WA 98007, USA }
\affiliation {Institute for Plasma Research, Bhat, Gandhinagar 382428, India }
\affiliation {The University of Sheffield, Sheffield S10 2TN, United Kingdom }
\affiliation {University of Southern California Information Sciences Institute, Marina Del Rey, CA 90292, USA }
\affiliation {California State University, Los Angeles, 5151 State University Dr, Los Angeles, CA 90032, USA }
\affiliation {Universit\`a di Trento, Dipartimento di Fisica, I-38123 Povo, Trento, Italy }
\affiliation {Montclair State University, Montclair, NJ 07043, USA }
\affiliation {National Astronomical Observatory of Japan, 2-21-1 Osawa, Mitaka, Tokyo 181-8588, Japan }
\affiliation {Canadian Institute for Theoretical Astrophysics, University of Toronto, Toronto, Ontario M5S 3H8, Canada }
\affiliation {Whitman College, 345 Boyer Avenue, Walla Walla, WA 99362 USA }
\affiliation {School of Mathematics, University of Edinburgh, Edinburgh EH9 3FD, United Kingdom }
\affiliation {University and Institute of Advanced Research, Gandhinagar Gujarat 382007, India }
\affiliation {IISER-TVM, CET Campus, Trivandrum Kerala 695016, India }
\affiliation {University of Szeged, D\'om t\'er 9, Szeged 6720, Hungary }
\affiliation {Tata Institute of Fundamental Research, Mumbai 400005, India }
\affiliation {INAF, Osservatorio Astronomico di Capodimonte, I-80131, Napoli, Italy }
\affiliation {University of Michigan, Ann Arbor, MI 48109, USA }
\affiliation {American University, Washington, D.C. 20016, USA }
\affiliation {Rochester Institute of Technology, Rochester, NY 14623, USA }
\affiliation {University of Bia{\l }ystok, 15-424 Bia{\l }ystok, Poland }
\affiliation {SUPA, University of Strathclyde, Glasgow G1 1XQ, United Kingdom }
\affiliation {University of Southampton, Southampton SO17 1BJ, United Kingdom }
\affiliation {University of Washington Bothell, 18115 Campus Way NE, Bothell, WA 98011, USA }
\affiliation {Institute of Applied Physics, Nizhny Novgorod, 603950, Russia }
\affiliation {Seoul National University, Seoul 08826, Korea }
\affiliation {Inje University Gimhae, South Gyeongsang 50834, Korea }
\affiliation {National Institute for Mathematical Sciences, Daejeon 34047, Korea }
\affiliation {NCBJ, 05-400 \'Swierk-Otwock, Poland }
\affiliation {Institute of Mathematics, Polish Academy of Sciences, 00656 Warsaw, Poland }
\affiliation {The School of Physics \& Astronomy, Monash University, Clayton 3800, Victoria, Australia }
\affiliation {Hanyang University, Seoul 04763, Korea }
\affiliation {NASA Marshall Space Flight Center, Huntsville, AL 35811, USA }
\affiliation {ESPCI, CNRS, F-75005 Paris, France }
\affiliation {Southern University and A\&M College, Baton Rouge, LA 70813, USA }
\affiliation {The University of Melbourne, Parkville, Victoria 3010, Australia }
\affiliation {College of William and Mary, Williamsburg, VA 23187, USA }
\affiliation {Indian Institute of Technology Madras, Chennai 600036, India }
\affiliation {IISER-Kolkata, Mohanpur, West Bengal 741252, India }
\affiliation {Scuola Normale Superiore, Piazza dei Cavalieri 7, I-56126 Pisa, Italy }
\affiliation {Universit\'e de Lyon, F-69361 Lyon, France }
\affiliation {Hobart and William Smith Colleges, Geneva, NY 14456, USA }
\affiliation {Janusz Gil Institute of Astronomy, University of Zielona G\'ora, 65-265 Zielona G\'ora, Poland }
\affiliation {University of Washington, Seattle, WA 98195, USA }
\affiliation {King's College London, University of London, London WC2R 2LS, United Kingdom }
\affiliation {Indian Institute of Technology, Gandhinagar Ahmedabad Gujarat 382424, India }
\affiliation {International Institute of Physics, Universidade Federal do Rio Grande do Norte, Natal RN 59078-970, Brazil }
\affiliation {Andrews University, Berrien Springs, MI 49104, USA }
\affiliation {Universit\`a di Siena, I-53100 Siena, Italy }
\affiliation {Trinity University, San Antonio, TX 78212, USA }
\affiliation {Abilene Christian University, Abilene, TX 79699, USA }
\affiliation {RMIT University, Melbourne, Victoria 3000, Australia}

\date{\today}

\begin{abstract}
	
	Results are presented from a semi-coherent search for continuous gravitational waves from the brightest low-mass X-ray binary, Scorpius X-1, using data collected during the first Advanced LIGO observing run (O1). The search combines a frequency domain matched filter (Bessel-weighted $\mathcal{F}$-statistic) with a hidden Markov model to track wandering of the neutron star spin frequency. No evidence of gravitational waves is found in the frequency range 60--650\,Hz. Frequentist 95\% confidence strain upper limits, $h_0^{95\%} = 4.0\times10^{-25}$, $8.3\times10^{-25}$, and $3.0\times10^{-25}$ for electromagnetically restricted source orientation, unknown polarization, and circular polarization, respectively, are reported at 106\,Hz. They are $\leq 10$ times higher than the theoretical torque-balance limit at 106\,Hz.
	
	\begin{description}
		\item[PACS numbers]
		95.85.Sz, 97.60.Jd
	\end{description}
\end{abstract}

\pacs{Valid PACS appear here}
\maketitle


\section{Introduction}

Rotating neutron stars are a possible source of persistent, periodic gravitational radiation. The signal is expected at specific multiples of the neutron star spin frequency $f_\star$ \cite{Riles2013}. Astrophysical models suggest that the radiation may be emitted at levels detectable by ground-based, long-baseline interferometers such as the Laser Interferometer Gravitational Wave Observatory (LIGO) and the Virgo detector \cite{aligo-Harry2010,aligo-Aasi2015,avirgo-Acernese2015,Andersson2011,Riles2013}. A time-varying quadrupole moment can result from thermal \cite{ushomirsky00, Johnson-McDaniel2013} or magnetic \cite{Cutler2002, Mastrano2011, Lasky2013} gradients, r-modes \cite{Owen1998, Heyl2002, Arras2002, bondarescu09}, or nonaxisymmetric circulation in the superfluid interior \cite{Peralta2006, VanEysden2008, Bennett2010, Melatos2015}.

Accreting neutron stars in binary systems are important search targets, because mass transfer spins up the star to $\gtrsim 10^2$\,Hz and may simultaneously drive several quadrupole-generating mechanisms \cite{Bildsten1998,Andersson1999,Nayyar2006,Melatos2007,Vigelius2009}. Moreover it is observed that the distribution of spin frequencies of low-mass X-ray binaries (LMXBs) cuts off near 620\,Hz \cite{Chakrabarty2003}, below the theoretical centrifugal break-up limit $\approx 1.4$\,kHz \cite{Cook1994}. This has been explained by hypothesizing that the gravitational radiation-reaction torque balances the accretion torque \cite{Papaloizou1978,Wagoner1984,Bildsten1998}, implying a relation between the X-ray flux and gravitational wave strain. Scorpius X-1 (Sco X-1), the most X-ray-luminous LMXB, is therefore a promising target for gravitational wave searches.

Initial LIGO achieved its design sensitivity over a wide band during LIGO Science Run 5 (S5) \cite{lscinstrument09} and exceeded it during Science Run 6 (S6) \cite{2012-S6}. The strain sensitivity of the next-generation Advanced LIGO interferometer is expected to improve ten-fold relative to Initial LIGO after several stages of upgrade \cite{Harry2010}. In the first observation run (O1), from September 2015 to January 2016, the strain noise is three to four times lower than in S6 across the most sensitive band, between 100\,Hz and 300\,Hz, and $\sim 30$ times lower around 50\,Hz \cite{Abbott2016-detector}.

Four types of searches have been conducted for Sco X-1 using data collected by Initial LIGO and Advanced LIGO (O1). None of these searches reported a detection.  First, a coherent search, using a maximum likelihood detection statistic called the $\mathcal{F}$-statistic \cite{Jaranowski1998}, analysed the most sensitive six-hour data segment from Science Run 2 (S2) and placed a 95\% confidence strain upper limit at $h_0^{95\%}\approx 2 \times 10^{-22}$ for two bands, 464--484\,Hz and 604--626\,Hz \cite{Abbott2007}. Second, a directed, semi-coherent analysis based on the sideband algorithm was conducted on a 10-day stretch of LIGO S5 data in the band 50--550\,Hz and reported median strain upper limits of $1.3 \times 10^{-24}$ and $8 \times 10^{-25}$ at 150\,Hz for arbitrary and electromagnetically restricted source orientations, respectively \cite{ScoX1-S5}. The sideband method sums incoherently the coherent $\mathcal{F}$-statistic power at frequency-modulated orbital sidebands and generates a new detection statistic called the $\mathcal{C}$-statistic \cite{Messenger2007,Sammut2014}. Third, a directed version of the all-sky TwoSpect search \cite{Goetz2011} was applied to S6 data and the second and third Virgo science runs (VSR2 and VSR3, respectively), yielding low-frequency upper limits of $h_0^{95\%}\approx 2 \times 10^{-23}$ in the band from 20\,Hz to 57.25\,Hz \cite{Aasi2014}. Another search of S6 data was carried out using the subsequently improved TwoSpect method \cite{Meadors2016}, spanning frequencies from 40\,Hz to 2040\,Hz and projected semi-major axis from 0.90\,s to 1.98\,s. It achieved a 95\% confidence level random-polarization upper limit of $h_0^{95\%} = 1.8 \times 10^{-24}$ at 165\,Hz \cite{Meadors2017}. Fourth, a directed version of the all-sky, radiometer search \cite{Ballmer2005} was conducted on all 20 days of Science Run 4 (S4) data \cite{Abbott2007-radiometer}, and was later applied to two years of S5 data, yielding a 90\% confidence root-mean-square strain upper limit of $7\times10^{-25}$ at 150\,Hz \cite{Abadie2011}, which converts to $h_0^{90\%}= 2 \times 10^{-24}$ \cite{MessengerNote}. The same method was applied to O1 data, yielding a median frequency-dependent limit of $h_0^{90\%}= 6.7 \times 10^{-25}$ at the most sensitive detector frequencies between 130--175\,Hz \cite{Radiometer_O1}.
 
It is probable that the spin frequency of Sco X-1 wanders stochastically under the fluctuating action of the hydromagnetic torque exerted by the accretion flow \cite{DeKool1993,Baykal1993,Bildsten1997}. Search methods that scan templates without guidance from a measured ephemeris are compromised because of spin wandering; for example the sideband search is restricted to a 10-day stretch of data in Ref. \cite{ScoX1-S5}, so that the signal power does not leak into adjacent frequency bins. Hidden Markov model (HMM) tracking offers a powerful strategy for detecting a spin-wandering signal \cite{Suvorova2016}. An HMM relates a sequence of observations to the most probable Markov sequence of allowed transitions between the states of an underlying, hidden state variable (here the gravitational wave signal frequency $f_0$) \cite{Quinn2001}. It can track $f_0$ over the total observation time $T_{\rm obs}$ by incoherently combining segments with duration $T_{\rm drift}=10$\,d of the output from a maximum-likelihood, coherent matched filter , improving the sensitivity by a factor $\approx (T_{\rm obs}/T_{\rm drift})^{1/4}$ relative to a single segment.

In this paper, we combine the sideband algorithm with an HMM and apply it to Advanced LIGO O1 data. Specifically we carry out a directed search for Sco X-1 in the band 60--650\,Hz. No evidence of a gravitational-wave signal is found. Frequentist 95\% confidence strain upper limits of $h_0^{95\%} = 4.0\times10^{-25}$, $8.3\times10^{-25}$, and $3.0\times10^{-25}$ are derived at 106\,Hz, for electromagnetically restricted source orientation, unknown polarization, and circular polarization, respectively. The paper is organized as follows. In Section II, we briefly review the search algorithm. In Section III, we discuss the astrophysical parameters of the source, search procedure, detection threshold and estimated sensitivity. Results of the search, including veto output, candidate follow-up, and gravitational wave strain upper limits are presented in Section IV. We discuss the torque-balance upper limit in Section V and conclude with a summary in Section VI.

\section{Method}
\label{sec:method}
In this section we briefly introduce the HMM formulation of frequency tracking and the Viterbi algorithm for solving the HMM in Section \ref{sec:HMMtracking} and Appendices \ref{sec:hmm} and \ref{sec:viterbi}, respectively. A matched filter appropriate for a continuous-wave source in a binary is reviewed in Section \ref{sec:matched_filter}.  A full description of the method can be found in Ref. \cite{Suvorova2016}.

\subsection{HMM tracking}
\label{sec:HMMtracking}
A HMM is a finite state automaton, in which a hidden (unobservable) state variable $q(t)$ transitions between values from the set $\{q_1, \cdots, q_{N_Q}\}$ at discrete times $\{t_0, \cdots, t_{N_T}\}$, while an observable state variable $o(t)$ transitions between values from the set $\{o_1, \cdots, o_{N_O}\}$. The probability that $q(t)$ jumps from state $q_i$ to state $q_j$ is given by the transition matrix $A_{q_i q_j}$. The likelihood that the hidden state $q_i$ gives rise to the observation $o_j$ is given by the emission probability $L_{o_j q_i}$. In this application, we map the discrete hidden states one-to-one to the frequency bins in the output of a frequency-domain estimator $G(f)$ (see Section \ref{sec:matched_filter}) computed over an interval of length $T_{\rm drift}$, with bin size $\Delta f_{\rm drift} = 1/(2 T_{\rm drift})$. The procedure for choosing $T_{\rm drift}$ is described in Appendix \ref{sec:hmm}.

For a Markov process, the probability that the hidden path $Q=\{q(t_0), \cdots, q(t_{N_T})\}$ gives rise to the observed sequence $O=\{o(t_0), \cdots, o(t_{N_T})\}$ is given by 
\begin{equation}
	\label{eqn:emi_prob}
	\begin{split}
		P(Q|O) = & L_{o(t_{N_T})q(t_{N_T})} A_{q(t_{N_T})q(t_{N_T-1})} \cdots L_{o(t_1)q(t_1)} \\ 
		& \times A_{q(t_1)q(t_0)} \Pi_{q(t_0)},
	\end{split}
\end{equation}
where $\Pi_{q_i}$ denotes the prior (see Appendix \ref{sec:hmm}). The classic Viterbi algorithm \cite{Viterbi1967} provides a recursive, computationally efficient route to computing $Q^*(O)$, the path that maximizes $P(Q|O)$. The steps in the algorithm are specified in Appendix \ref{sec:viterbi}; the number of operations is of order $(N_T+1)N_Q \ln N_Q$ \cite{Quinn2001}. In this paper, we define a detection score $S$, such that the log likelihood of the optimal Viterbi path equals the mean log likelihood of all paths plus $S$ standard deviations, viz.
\begin{equation}
	S = \frac{\ln \delta_{q^*}{(t_{N_T})} -\mu_{\ln \delta}(t_{N_T})}{\sigma_{\ln \delta}(t_{N_T})}
\end{equation}
with
\begin{equation}
	\mu_{\ln \delta}(t_{N_T}) = N_Q^{-1} \sum_{i=1}^{N_Q} \ln \delta_{q_i}(t_{N_T})
\end{equation}
and
\begin{equation}
	\sigma_{\ln \delta}(t_{N_T})^2 = N_Q^{-1} \sum_{i=1}^{N_Q} [\ln \delta_{q_i}(t_{N_T}) - \mu_{\ln \delta}(t_{N_T}) ]^2,
\end{equation}
where $\delta_{q_i}(t_{N_T})$ denotes the maximum probability of the path ending in state $q_i$ ($1\leq i \leq N_Q$) at step $N_T$ (see Appendix \ref{sec:viterbi}), and $\delta_{q^*}{(t_{N_T})}$ is the likelihood of the optimal Viterbi path, i.e. $P[Q^*(O)|O]$.

\subsection{Matched filter: Bessel-weighted $\mathcal{F}$-statistic}
\label{sec:matched_filter}
The emission probability $L_{o(t) q_i}$ is computed from a frequency-domain estimator $G(f)$ as described in Appendix \ref{sec:hmm}. In the context of continuous-wave searches, $G(f)$ is a matched filter. The optimal matched filter for a biaxial rotor with no orbital motion is the maximum-likelihood $\mathcal{F}$-statistic \cite{Jaranowski1998}, which accounts for the rotation of the Earth and its orbit around the Solar System barycenter (SSB). When the source orbits a binary companion, the gravitational-wave signal frequency is modulated due to the orbital Doppler effect \cite{Ransom2003,Messenger2007,Sammut2014}. The $\mathcal{F}$-statistic power is distributed into approximately $M=2m+1$ orbital sidebands with $m = \text{ceil} (2 \pi f_0 a_0)$, separated in frequency by $1/P$, where $f_0$ is the intrinsic gravitational wave frequency, $a_0$ is the light travel time across the projected semi-major axis of the orbit, $P$ is the orbital period, and $\text{ceil}(x)$ denotes the smallest integer greater than or equal to $x$. For a Keplerian orbit with zero eccentricity, the gravitational wave strain can be expanded in a Jacobi-Anger series as \cite{Abramowitz1964,Suvorova2016}
\begin{equation}
	\label{eqn:wave_strain_expansion}
	h(t) \propto \mathop{\sum} \limits_{n=-\infty}^{\infty} J_n(2 \pi f_0 a_0) \cos [2\pi (f_0 + n/P)t],
\end{equation}
where $J_n(z)$ is a Bessel function of order $n$ of the first kind. The mathematical form of (\ref{eqn:wave_strain_expansion}) suggests a Bessel-weighted $\mathcal{F}$-statistic as the matched filter $G(f)$ for a biaxial rotor in a binary system, which can be expressed as the convolution \cite{Suvorova2016}
\begin{equation}
	\label{eqn:matched_filter}
	G(f)=\mathcal{F}(f) \otimes B(f),
\end{equation}
where $B(f)$ is given by
\begin{equation}
	\label{eqn:matched_filter_cont}
	B(f)=\sum\limits_{n=-(M-1)/2}^{(M-1)/2} [J_n(2 \pi f a_0)]^2 \delta(f-n/P).
\end{equation}

Compared to the $\mathcal{C}$-statistic, used in a previously published sideband search for Sco X-1 \cite{Sammut2014,Suvorova2016}, where the factor $[J_n(2 \pi f a_0)]^2$ in (\ref{eqn:matched_filter_cont}) is replaced by unity, the Bessel-weighted matched filter recovers approximately $\sqrt{2}$ times more signal. It marshals more power into a single bin, producing a distinct spike with shoulders, instead of the relatively flat onion-dome peak produced by the $\mathcal{C}$-statistic. These characteristics facilitate Viterbi tracking (see Section IV A in Ref. \cite{Suvorova2016} for details). We leverage the existing, efficient, thoroughly tested $\mathcal{F}$-statistic software infrastructure in the LSC Algorithm Library Applications (LALApps)\footnote{https://www.lsc-group.phys.uwm.edu/daswg/projects/lalapps/} to compute $\mathcal{F}(f)$ in (\ref{eqn:matched_filter}) \cite{F-stat2011}.

\section{Implementation}
\label{sec:implementation}
In this section we introduce the electromagnetically measured source parameters of Sco X-1 (Section \ref{sec:scox1-para}), and describe the workflow of the pipeline (Section \ref{sec:workflow}), detection threshold (Section \ref{sec:threshold}), and search sensitivity (Section \ref{sec:sensitivity}).

\subsection{Sco X-1 parameters}
\label{sec:scox1-para}
The sky position ($\alpha, \delta$), orbital elements ($a_0, P$), and orientation angles ($\iota, \psi$) of Sco X-1 have been measured electromagnetically to various degrees of accuracy. The values and $1\sigma$ (68\%) confidence level uncertainties are quoted in the top half of Table \ref{tab:paras}. 

The published uncertainty in the orbital period, $\Delta P = 0.0432$\,s \cite{Galloway2014}, restricts the coherent observation time to $T_{\rm drift} \leq 50$\,d \cite{Sammut2014,ScoX1-S5}. Hence it is safe to take a single, fixed $P$ value when evaluating the $\mathcal{F}$-statistic, given that the coherent data stretches we analyse are limited to 10\,d (20\,d for follow up; see Section \ref{sec:longer_coherent_time_followup}). The published uncertainty in the projected semi-major axis, inferred from the measured orbital velocity, is $\Delta a_0 = 0.18$\,s \cite{Steeghs2002}. In the previous S5 sideband search, it is demonstrated that taking a single, fixed $a_0$ value does not impact search sensitivity given this published uncertainty \cite{Sammut2014,ScoX1-S5}. However recent, unpublished research has revised the range of $a_0$ upwards to $0.36\,{\rm s} \leq a_0 \leq 3.25$\,s. This is because the orbital velocity is difficult to measure electromagnetically, and the previous measurement is based on searching for the optimal centre of symmetry in the accretion disk emission, yielding an estimated velocity of $40\pm5$\,km\,s$^{-1}$ \cite{Steeghs2002}. The preliminary results from the more recent study, which uses Doppler tomography measurements and Markov Chain Monte-Carlo analysis for the velocity, show that the constraint on the orbital velocity is weaker, corresponding to a range from $10\,{\rm km}\,{\rm s}^{-1}$ to $90\,{\rm km}\,{\rm s}^{-1}$ \cite{Wang_thesis,Wang_comm}. It is shown in Section IV B of Ref. \cite{Suvorova2016} that, if the true value of $a_0$ differs from the estimated $a_0$ by 10\%, it would produce an uncertainty in the estimated frequency of $\approx 0.001$\,Hz. Moreover, the log likelihood of the optimal path decreases by $\sim 50\%$, if the true value of $a_0$ differs from the estimated $a_0$ by 25\%. We search over the wider, unpublished range of $a_0$ with a resolution of 0.01805\,s in order to preserve sensitivity. The orientation angles $\iota$ and $\psi$  are measured from the position angle of the Sco X-1 radio jets on the sky, assuming  that the rotation axis of the neutron star is perpendicular to the accretion disk. In the previously published sideband search, two orientation priors are considered: (1) uniform distributions of $\cos \iota$ and $\psi$; and (2) distributions peaked around the observed values in the top half of Table \ref{tab:paras}. 

The parameter space covered by the search is defined in the bottom half of Table \ref{tab:paras}. We assume uniform priors on both $f_0$ and $a_0$. 

\begin{table*}[!tbh]
	\centering
	\setlength{\tabcolsep}{7pt}
	\renewcommand\arraystretch{1.1}
	\begin{tabular}{llll}
		\hline
		\hline
		Observed parameter & Symbol & Value & Reference \\
		\hline
		Right ascension & $\alpha$ & 16h 19m 55.0850s &\cite{Bradshaw1999} \\
		Declination & $\delta$ & $-15^{\circ}38' 24.9''$ &\cite{Bradshaw1999}\\
		X-ray flux & $F_X$ & $4\times10^{-7}\,{\rm erg}\,{\rm cm}^{-2}\,{\rm s}^{-1}$& \cite{Watts2008}\\
		Orbital period & $P$ & $68023.70496\pm 0.0432$\,s &\cite{Galloway2014}\\
		Projected semi-major axis& $a_0$& $1.44\pm 0.18$\,s &\cite{Steeghs2002}\\
		Polarization angle &$\psi$& $234\pm 3^{\circ}$&\cite{Fomalont2001}\\
		Inclination angle & $\iota$ & $44\pm 6^{\circ}$&\cite{Fomalont2001}\\
		\hline
		Search parameter & Symbol & Search range & Resolution \\
		\hline
		Frequency &$f_0$ & $60-650$\,Hz & $5.787037 \times 10^{-7}$\,Hz \\
		Projected semi-major axis & $a_0$& $0.361-3.249$\,s &0.01805\,s \\
		\hline
		\hline
	\end{tabular}
	\caption{Electromagnetically observed parameters (top half) and search parameters (bottom half) for Sco X-1. The uncertainties are at the $1\sigma$ confidence level.}
	\label{tab:paras}
\end{table*}

\subsection{Workflow}
\label{sec:workflow}
The search is parallelized into 1-Hz sub-bands to assist with managing the relatively large volume of data involved. The sub-bands must be narrow enough, so that we can replace $f$ with the mean value $\bar{f}$ in each sub-band to a good approximation, in order to avoid recalculating $B(f)$ in every frequency bin. The sub-bands must also be wide enough to contain the width of the matched filter. Sub-bands of 1-Hz satisfy both of these requirements, and were also adopted in the S5 sideband search \cite{ScoX1-S5}.

\begin{figure}[!tbh]
	\centering
	\scalebox{0.53}{\includegraphics{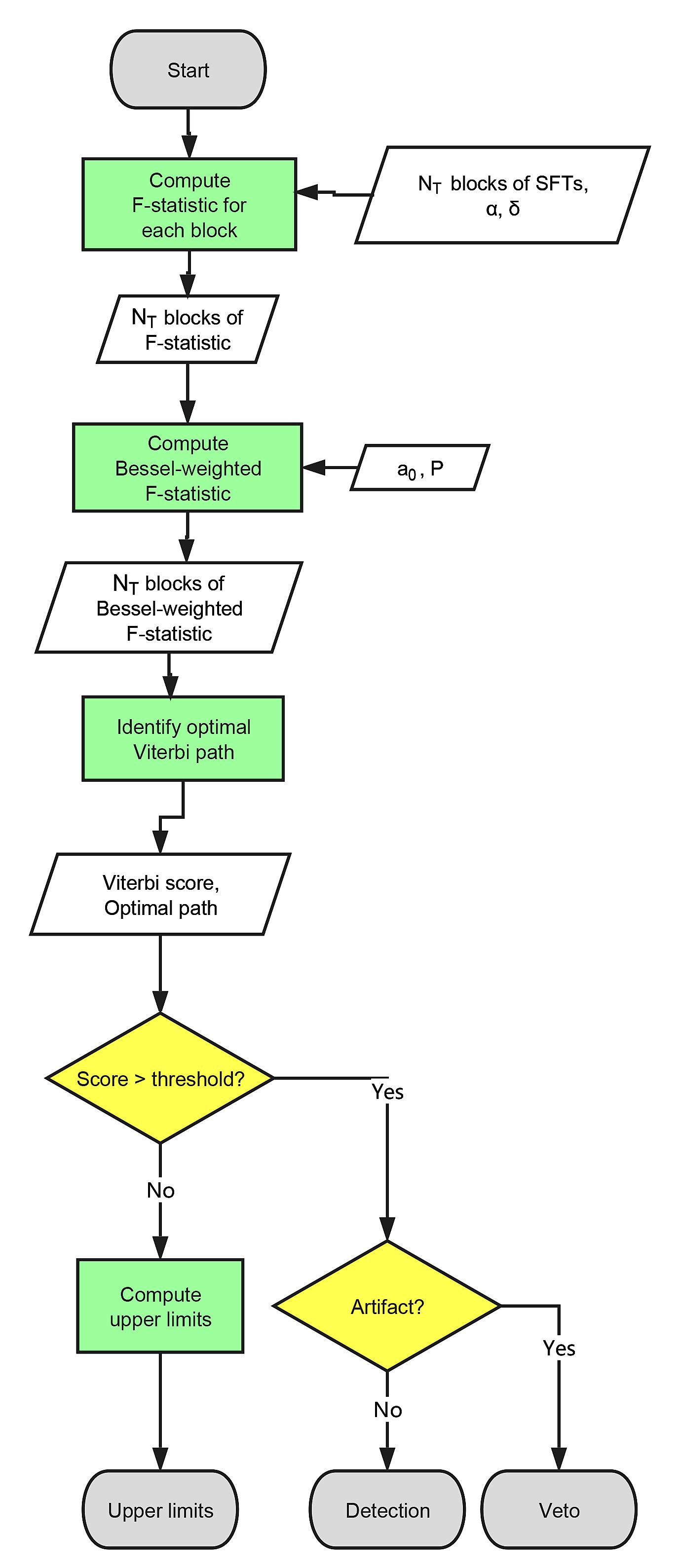}}
	\caption{Flowchart of the pipeline in each 1-Hz sub-band.}
	\label{fig:flowchart}
\end{figure}

The flow chart in Figure \ref{fig:flowchart} summarises the procedural steps in the search pipeline. Firstly, the 30-min short Fourier transforms (SFTs) constituting the whole observation are divided into $N_T$ blocks, each of duration $T_{\rm drift}=10$\,d. In each 1-Hz sub-band, the $\mathcal{F}$-statistic is computed for each block at the known sky location of the source. Next we compute the Bessel-weighted $\mathcal{F}$-statistic $G(f)$ from (\ref{eqn:matched_filter}) and (\ref{eqn:matched_filter_cont}), taking $a_0$ and $P$ as inputs; that is, $G(f)$ is computed in $N_{f_0}$ frequency bins for each of the $N_T$ blocks. Theoretically the HMM hidden state variable is two-dimensional, because we search over $f_0$ and $a_0$. In practice $a_0$ varies imperceptibly on the time-scale $T_{\rm obs}$, so the algorithm is equivalent to multiple, independent, one-dimensional HMM searches over $f_0$ on a grid of $a_0$ values. The detection score and corresponding optimal Viterbi path are recorded in each 1-Hz sub-band. We evaluate the detection scores to identify candidates, judge whether or not they come from instrumental artifacts via a well-defined hierarchy of vetoes, and claim a detection or compute strain upper limits for sub-bands without candidates.

\subsection{Threshold}
\label{sec:threshold}
We determine the Viterbi score threshold $S_{\rm th}$ for a given false alarm rate $\alpha_{\rm f}$ through Monte-Carlo simulations, such that searching data sets containing pure noise yields a fraction $\alpha_{\rm f}$ of positive detections with $S>S_{\rm th}$. SFTs containing pure Gaussian noise are generated for seven 1-Hz sub-bands, starting at 55\,Hz, 155\,Hz, 255\,Hz, 355\,Hz, 455\,Hz, 555\,Hz, and 650\,Hz, with the same single-sided power spectral density (PSD) $S_h (f)$ as actual O1 data and with $T_{\rm obs}=130$\,d. Searches are repeated for 100 noise realisations in each 1-Hz sub-band following the recipe in Fig. \ref{fig:flowchart}. We track 161 $a_0$ values from 0.361\,s to 3.249\,s, with resolution 0.01805\,s, as for a real search. We find that the results depend weakly on the sub-bands: the mean $\langle S \rangle$ varies from 6.48 to 6.59, and the standard deviation $\sigma_S$ varies from 0.24 to 0.33. Combining the 700 realisations yields $S_\text{th} = 7.34$ for $\alpha_{\rm f}=1\%$. 

To check the influence of non-Gaussian noise on $S_{\rm th}$, we choose three 1-Hz sub-bands, starting at 157\,Hz, 355\,Hz, and 635\,Hz, in O1 interferometer data and repeat the search for real noise. As we have no means of generating multiple, random, real-noise realisations from scratch, we take 100 different sky locations as background noise realisations. We find that $\langle S \rangle$ and $\sigma_S$ range from 6.36 to 6.38 and 0.27 to 0.34, respectively. These results match the output from Gaussian noise simulations to better than $\sim 3\%$, as does $S_{\rm th}$. Hence we set $S_\text{th} = 7.34$ in the forthcoming analysis described in Section \ref{sec:O1_search}.

In the follow-up procedures in Section \ref{sec:O1_search}, we search a subset of the data either from a single interferometer with $T_{\rm obs}=130\,{\rm d} = 13\,T_{\rm drift}$ or two interferometers with $T_{\rm obs}=60\,{\rm d} = 6\,T_{\rm drift}$. To check the validity of $S_{\rm th}=7.34$ when searching a subset of the data, we run 400 trials of Gaussian noise simulations using data generated for a single interferometer with $T_{\rm obs}=130\,{\rm d}$ or two interferometers with $T_{\rm obs}=60\,{\rm d}$. The resulting $S_{\rm th}$ remains the same overall, and $\langle S \rangle$ and $\sigma_S$ range from 6.44 to 6.50 and 0.27 to 0.30, respectively, matching the output in the simulations with two interferometers and $T_{\rm obs}=130\,{\rm d}$ to better than $\sim 3\%$. Hence we keep $S_\text{th} = 7.34$ fixed for the follow-up procedures in Section \ref{sec:O1_search}.

\subsection{Sensitivity}
\label{sec:sensitivity}
\begin{figure*}
	\centering
	\scalebox{0.4}{\includegraphics{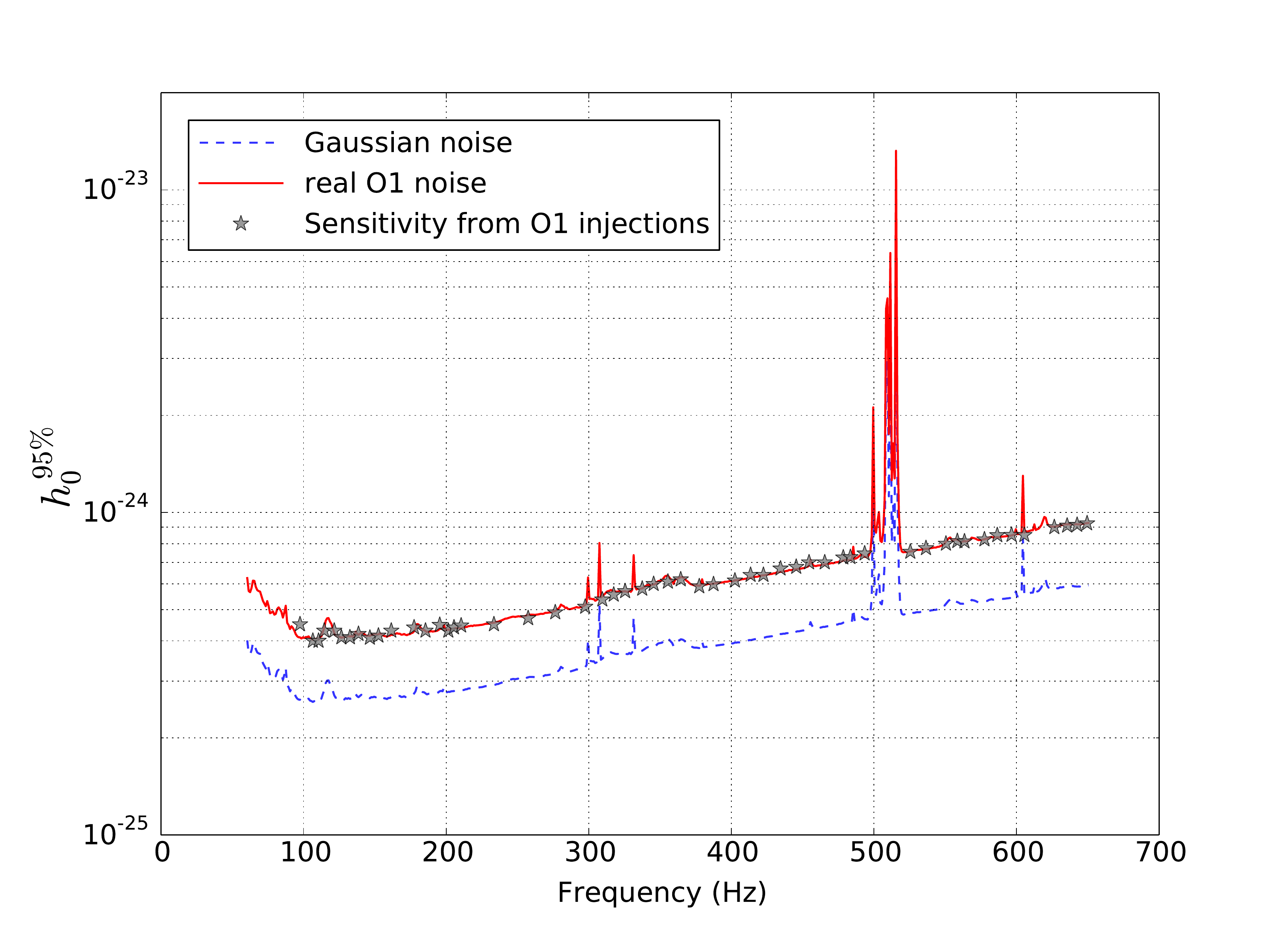}}
	\caption{Characteristic wave strain for 95\% detection efficiency, $h_0^{95\%}$, versus frequency (Hz) from Monte-Carlo simulations. Signals are injected with a restricted inclination angle $\cos\iota_{\rm inj} \approx 0.7193$. Blue dashed curve: $h_0^{95\%}$ from Gaussian noise with $S_h(f)$ evaluated from the nominal O1 PSD and $T_{\rm obs}=130$\,d. Gray stars: $h_0^{95\%}$ from injections into real O1 interferometer data in 53 1-Hz sub-bands. Red solid curve: $h_0^{95\%}$ in real O1 noise, corrected for duty cycle and nongaussianity by multiplying the blue dashed curve by a factor $\kappa=1.56$. The red solid curve overlaps substantially with the gray stars.}
	\label{fig:sensitivity}
\end{figure*}

Given the threshold $S_{\rm th}\,(\alpha_{\rm f}=1\%)=7.34$, we evaluate the characteristic wave strain yielding 95\% detection efficiency (i.e. 5\% false dismissal rate), denoted by $h_0^{95\%}$, through Monte-Carlo simulations with signals injected into Gaussian noise. The simulations are performed between 155--156\,Hz, where the detectors are most sensitive, with $T_{\rm obs}=130$\,d, $T_{\rm drift}=10$\,d, $N_T=13$, $\sqrt{S_h}=1 \times 10^{-23}\,{\rm Hz}^{-1/2}$, and source parameters copied from Table \ref{tab:paras}. We choose $T_{\rm obs}=130$\,d to equal the duration of O1. The parameters ${f_0}_{\rm inj}$, ${a_0}_{\rm inj}$, $\cos \iota_{\rm inj}$, and $\psi_{\rm inj}$ are randomly chosen with a uniform distribution within the ranges 155.34565530--155.3456847\,Hz, 0.36--3.25\,s, 0.712107--0.726493, and 0--$2\pi$\,rad, respectively. We obtain $h_0 ^{95\%} = 3\times 10^{-25}$ for electromagnetically restricted orientation by assuming $\iota \approx 44^\circ$ \cite{Fomalont2001}. In reality, the signal-to-noise ratio scales in proportion to $h_0^{\rm eff}$, given by 
\begin{equation}
	\label{eqn:scaling_ul}
	h_0^{\rm eff}= h_0\,2^{-1/2}\{{[(1+\cos^2\iota)/2]^2 + \cos^2\iota}\}^{1/2},
\end{equation}
rather than $h_0$ \cite{Jaranowski1998,Messenger2015}. Hence we can convert the limiting wave strain to $h_0^{{\rm eff},95\%} \approx 0.74 h_0^{95\%}$ using the value $\iota=44^\circ$. For $T_{\rm obs}$ fixed, we expect 
\begin{equation}
	\label{eqn:sensi_curve}
	h_0^{95\%} \propto S_h^{1/2}f_0^{1/4}.
\end{equation}
The latter scaling is verified by a group of injections in three other frequency bands (55--56\,Hz, 355--356\,Hz, and 649--650\,Hz). Evaluating $S_h(f)$ from the O1 PSD, we plot $h_0^{95\%}$ versus $f_0$ as the blue dashed curve in Figure~\ref{fig:sensitivity}, which represents the 95\% detection efficiency curve in Gaussian noise simulations.

In practice interferometer noise is non-Gaussian, and $T_{\rm obs}$ is less than 130\,d (duty cycle $\approx 60\%$). To correct for this, we pick 53 1-Hz sub-bands, run 3000 injections in real O1 interferometer data, and compare the resulting $h_0^{95\%}$ to the blue dashed curve in Figure~\ref{fig:sensitivity}. The injected signal parameters are chosen in the same way as in the Gaussian noise simulation. In each sub-band tested, the resulting $h_0^{95\%}$ values from real O1 injections are plotted as gray stars in Figure~\ref{fig:sensitivity}. The correction factor $\kappa$ in each 1-Hz sub-band is defined as $h_0^{95\%}$, as marked by the gray star, divided by the value read off the blue dashed curve. The correction factors in 53 sub-bands fluctuate weakly, with mean $\langle \kappa \rangle=1.56$ and standard deviation $\sigma_\kappa = 0.03$. We therefore apply the same $\kappa=1.56$ across the full search and adjust the blue dashed curve to give the red solid curve in Figure \ref{fig:sensitivity}. The latter represents the characteristic wave strain for 95\% detection efficiency as a function of frequency in real O1 data. We find that 2846 out of the 3000 O1 injections are detected with $S>S_{\rm th}$, yielding a detection rate of 94.87\%, consistent with the targeted detection efficiency.

\section{O1 analysis}
\label{sec:O1_search}
In this section, we analyse data from the O1 observing run extending from 12 September 2015 to 19 January 2016 UTC (GPS time 1126051217 to 1137254417). The data are divided into 13 blocks, with $T_{\rm drift}=10$\,d, and fed into the HMM tracker described in Section \ref{sec:method} and \ref{sec:implementation}.

Narrowband, instrumental noise lines (e.g. power line at 60\,Hz, beam splitter violin mode, electronics, mirror suspension, calibration) and their harmonics can obscure astrophysical continuous-wave signals. At low frequencies between 25\,Hz and 60\,Hz, there are at least six known lines in each 1-Hz sub-band, and $\approx 2/3$ of the sub-bands contain more than 15 lines. Hence we do not search below 60\,Hz, because the optimal paths returned by the HMM are dominated by difficult-to-model noise. The sensitivity of the method degrades, as the width $4\pi a_0 f_0/P$ of the matched filter increases (see Section \ref{sec:matched_filter}). We terminate the search arbitrarily at $f_0=650$\,Hz to keep $4\pi a_0 f_0/P$ below $\approx 0.4$\,Hz, which is almost half the width of a sub-band.

We record the first-pass candidates identified by the search in Figure \ref{fig:candidates}. We then sift them through a systematic hierachy of vetoes as follows: (1) known instrumental line veto (Section \ref{sec:line_veto}), (2) single interferometer veto (Section \ref{sec:ifo_veto}), (3) $T_{\rm obs}/2$ veto (Section \ref{sec:half_veto}), and (4) $T_{\rm drift}$ veto (Section \ref{sec:longer_coherent_time_followup}). The safety verification of the four-step veto procedure is described in Section \ref{sec:veto_safety}. Table \ref{tab:outlier_num} lists the numbers of candidates surviving after each veto. No candidate survives all the vetoes and so we set upper limits on $h_0$. The strain upper limits are discussed in Section \ref{sec:upper_limit}.

\begin{table}
	\centering
	\setlength{\tabcolsep}{7pt}
	\renewcommand\arraystretch{1.1}
	\begin{tabular}{ll}
		\hline
		\hline
		Veto & Number\\
		\hline
		First pass & 180  \\
		After line veto & 44  \\
		After single IFO veto & 6  \\
		After half $T_{\rm obs}$ veto & 2 \\
		After longer $T_{\rm drift}$ veto & 0\\
		\hline
	\end{tabular}
	\caption{Number of candidates surviving each veto.}
	\label{tab:outlier_num}
\end{table}

\subsection{Vetoes}
\subsubsection{Known line veto}
\label{sec:line_veto}
\begin{figure*}
	\centering
	\scalebox{0.21}{\includegraphics{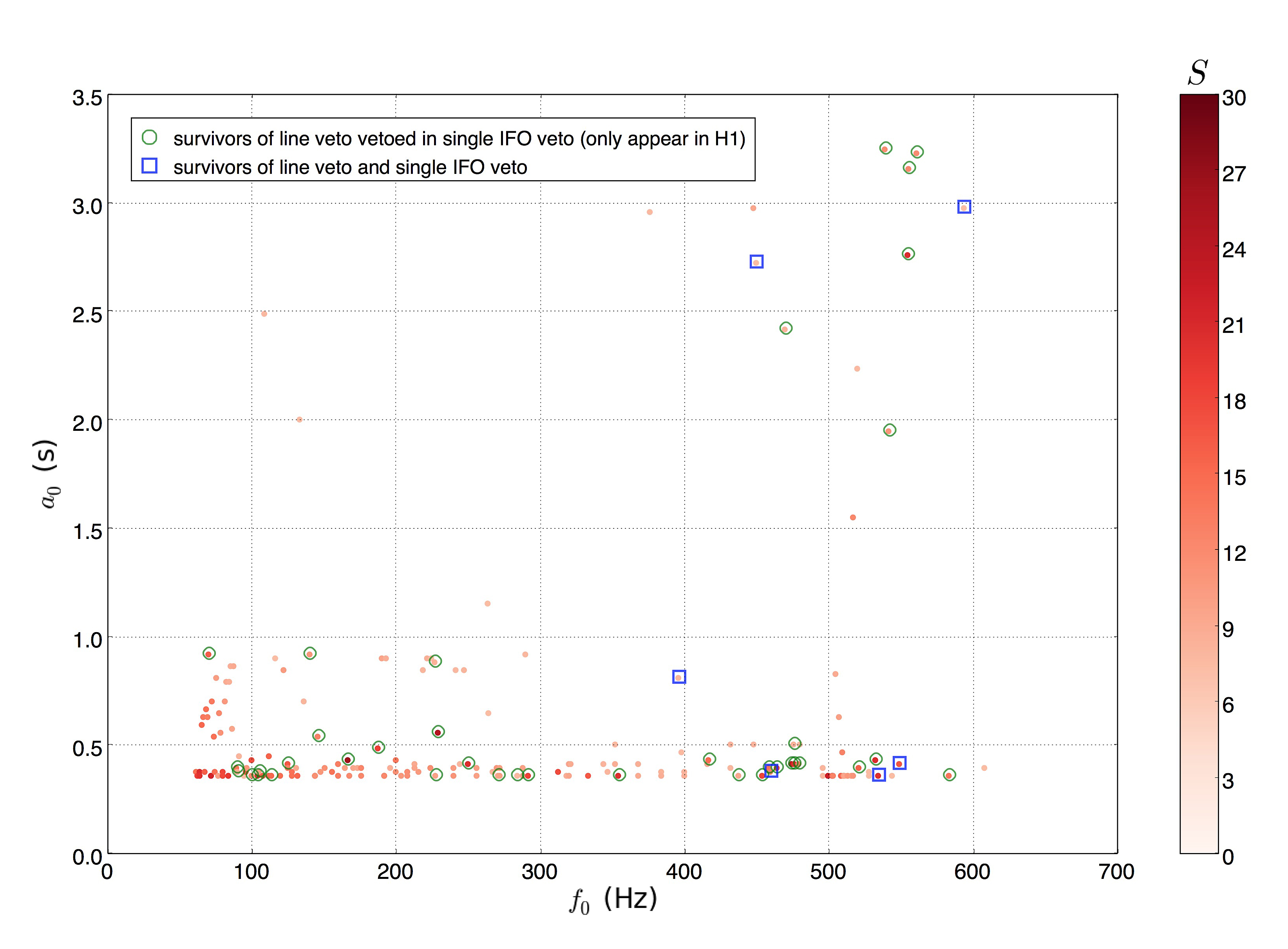}}
	\caption{First-pass candidates and survivors of the known line veto and single interferometer veto. The detection score $S$ in each 1-Hz sub-band is plotted as a function of $f_0$ and $a_0$ as estimated by the HMM. Each red dot stands for one candidate with $S > S_{\rm th}=7.34$. The colour of the dots scales with $S$ (see colour bar at right). Red dots without green circles or blue squares are vetoed due to contamination by known instrumental lines. Candidates are marked by green circles if they are detected with higher $S$ in H1 than the original score but not detected in L1. Green circles are vetoed (category A in Table \ref{tab:single_ifo}). None of the candidates is detected with higher $S$ in L1 than the original score while not being detected in H1. Candidates marked by blue squares survive both the known line veto and the single interferometer veto and require further follow-up.}
	\label{fig:candidates}
\end{figure*}

First-pass candidates with $S>S_{\rm th}=7.34$ (red dots) are plotted in Figure \ref{fig:candidates} as a function of $f_0$ and $a_0$ as estimated by the HMM. Each dot stands for a candidate in a 1-Hz sub-band. The colour of a dot indicates its associated $S$ value (higher $S$ in darker shade). The HMM returns an optimal path $f_0(t)$ whose wandering is too slight to be discerned visually in Figure \ref{fig:candidates}. We take $f_0$ to equal the arithmetic mean of the min $f_0(t)$ and max $f_0(t)$ in the plot. 

A candidate is vetoed, if $f_0(t)$ satisfies $|f_0(t)-f_{\rm line}| < 4\pi a_0 f_0/P$ \emph{anywhere} on the path, where $f_{\rm line}$ is the frequency of a known instrumental noise line. We find that the line veto excludes 75\% of the candidates. The 44 survivors are marked by green circles or blue squares in Figure \ref{fig:candidates}. (The distinction between the green and blue symbols is discussed below.) One immediately notices that most of the red dots appear at $a_0 \lesssim 0.5$\,s for all $f_0$. This is because a narrower matched filter produces a higher score when it encounters a narrow noise line. A noise line that produces high $\mathcal{F}$-statistic values concentrated in a handful of frequency bins spreads out when convolved with the matched filter in (\ref{eqn:matched_filter_cont}) and contributes to every Bessel-weighted $\mathcal{F}$-statistic bin in the band $|f_0(t)-f_{\rm line}| < 4\pi a_0 f_0/P$. The Viterbi score computed from the log likelihood of the optimal path is normalized by the standard deviation of all the log likelihoods in a 1-Hz sub-band. It is higher if the $\mathcal{F}$-statistic output containing a noise line is convolved with a narrower matched filter (i.e. smaller $a_0$), because the $\mathcal{F}$-statistic-processed noise-line power is dispersed into fewer orbital sidebands. The plot confirms that most vetoed candidates have $a_0 \lesssim 0.5$\,s.

Instrumental lines are picked up readily by the HMM, rendering any astrophysical signal invisible in the relevant 1-Hz sub-band. One might seek to improve the search by notching out the instrumental lines first, before applying the HMM to the rest of the sub-band. However, O1 lines cluster closely below 90\,Hz and near 300\,Hz and 500\,Hz, fragmenting the uncontaminated bands. It is onerous to circumvent the fragmentation, so we postpone this improvement to future searches, when better interferometer sensitivity will warrant the extra effort. In this search, we do not report results in a 1-Hz sub-band, if the optimal path intersects any instrumental line. In total 136 out of 591 1-Hz sub-bands are removed in this way.

\subsubsection{Single interferometer veto}
\label{sec:ifo_veto}
We now examine the 44 candidates surviving the known line veto by searching data from H1 and L1 separately. The sensitivities of the two interferometers during O1 are comparable, implying either that an astrophysical signal should appear in both detectors if it is strong enough or that it cannot be detected in either detector but can be seen after combining data from both. In contrast, a candidate is more likely a noise artifact originating in a single detector if it is detected in one detector with higher $S$ than the original combined score $S_\cup$, while the other detector yields $S<S_{\rm th}$. 

We can categorize survivors of the known line veto in Section \ref{sec:line_veto} into four classes presented in Table \ref{tab:single_ifo}. 

\emph{Category A}: Only one detector yields $S > S_{\rm th}$, equal to or higher than $S_\cup$; and the frequency estimated from the detector with $S \geq S_\cup$ is approximately equal to that obtained by combining both, with an absolute discrepancy less than $2 \pi {a_0}_\cup {f_0}_\cup/P$, where ${a_0}_\cup$ and ${f_0}_\cup$ are the $a_0$ and $f_0$ estimated using both detectors. Typically we find that the absolute discrepancy is less than 0.01\,Hz, even smaller than $2 \pi {a_0}_\cup {f_0}_\cup/P$. Any astrophysical signal that is too weak to yield $S> S_{\rm th}$ in one detector is unavoidably obscured by the undocumented noise artifact in the other detector. Hence we veto candidates in category A.

\emph{Category B}: Only one detector yields $S > S_{\rm th}$, equal to or higher than $S_\cup$, but the optimal path from the detector with $S \geq S_\cup$ occurs at $f_0$ with $|f_0 -{f_0}_\cup| \geq 2 \pi {a_0}_\cup {f_0}_\cup/P$ (denoted by $f_0 \neq {f_0}_\cup$ in Table \ref{tab:single_ifo}). It is possible that a real signal only shows up at ${f_0}_\cup$ after combining data from two detectors. Hence we keep candidates in category B for follow-up.

\emph{Category C}: Both detectors yields $S \geq S_{\rm th}$. The candidate may come either from noise or from a real signal registering strongly in both detectors. Hence we keep candidates in category C for follow-up. 

\emph{Category D}: Both detectors yield $S < S_{\rm th}$ even though we have $S_\cup \geq S_{\rm th}$. A real signal may be too weak to register in either detector individually but rises above the noise, when the two detectors are combined. Hence we keep candidates in categories D for further examination.

\begin{table*}
	\centering
	\setlength{\tabcolsep}{6pt}
	\renewcommand\arraystretch{1.1}
	\begin{tabular}{llll}
		\hline
		\hline
		Category &Score in one detector $S$ & Estimated frequency in one detector $f_0$ & Action\\
		\hline
		A&$S \geq S_\cup$ in one detector but  $S < S_{\rm th}$ in the other & $f_0 \approx {f_0}_\cup$ where $S \geq S_\cup$ & Veto\\
		B&$S \geq S_\cup$ in one detector but  $S < S_{\rm th}$ in the other & $f_0 \neq {f_0}_\cup$ where $S \geq S_\cup$ & Keep\\
		C&$S \geq S_{\rm th}$ in both detectors &  & Keep\\
		D&$S < S_{\rm th}$ in both detectors &  & Keep\\
		\hline
		\hline
	\end{tabular}
	\caption{Actions to be taken for survivors of the known line veto in Section \ref{sec:line_veto} according to the score $S$ and the estimated frequency $f_0$ from each single detector. $S_\cup$ and ${f_0}_\cup$ stand for the score and estimated frequency yielded by the original search combining two detectors.}
	\label{tab:single_ifo}
\end{table*}

Among the 44 candidates surviving the line veto, 38 in total are vetoed. They are marked by green circles in Figure \ref{fig:candidates}. All of them only appear in H1. The remaining six candidates marked by blue squares need to be examined further manually. Four of them show higher scores in H1 and $S<S_{\rm th}$ in L1, but the estimated $f_0$ from H1 is different from that obtained by combining both detectors, falling into category B in Table \ref{tab:single_ifo}. Two candidates, in the sub-bands 449--450\,Hz and 593--594\,Hz, fall into category D in Table \ref{tab:single_ifo}, with $S < S_{\rm th}$ in both H1 and L1.

\subsubsection{$T_{\rm obs}/2$ veto}
\label{sec:half_veto}
We now divide the observing run into two halves: 12 September 2015 to 20 November 2015 UTC (GPS time 1126051217 to 1132020365) and 20 November 2015 to 19 January 2016 UTC (GPS time 1132020366 to 1137254417). We search the halves separately in the six 1-Hz sub-bands containing veto survivors listed in Table \ref{tab:candidates_veto2}, combining data from two interferometers. Similar to the criteria listed in Section \ref{sec:ifo_veto}, we veto a candidate, if it appears in one half, with $S \geq S_\cup$, but does not appear in the other half, and if the estimated $f_0$ value is approximately equal to the original value.

The three candidates near 459\,Hz, 534\,Hz, and 548\,Hz appear in the first half with higher $S$ but not in the second half. The candidate near 395\,Hz appears in the second half with higher $S$ but not in the first half. Each one of them is detected in the first or second half at a frequency approximately equal to the original estimated $f_0$ with absolute discrepancy less than 0.01\,Hz.

In sub-bands 449\,Hz and 593\,Hz, neither of the two halves yields $S>S_{\rm th}$. These two candidates are marked by an asterisk in Table \ref{tab:candidates_veto2} and require further follow-up.

\begin{table*}
	\centering
	\setlength{\tabcolsep}{7pt}
	\renewcommand\arraystretch{1.1}
	\begin{tabular}{lllllll}
		\hline
		\hline
		Sub-band (Hz) & $f_0$ (Hz) & $f_{\rm max} - f_{\rm min}$ & $a_0$ (s) & $S_\cup$& $S_{\rm 1st\, half}$ & $S_{\rm 2nd\,half}$ \\
		&   & ($\Delta f_{\rm drift}$) & & & &  \\
		\hline
		395--396 &   395.8561536 & 3 & 0.81 &8.05153 & 6.55545 & 9.13679 \\
		449--450* & 449.8116935 & 3 & 2.73 & 7.38701 & 6.46122 & 6.50190\\
		459--460 &   459.5557459  & 6 & 0.38 & 12.76130 & 14.61070 & 6.30887 \\
		534--535 &   534.3625717   & 4 & 0.36 &20.18630 & 20.53770 & 6.97788 \\
		548--549 &   548.9457104  & 7 & 0.42 & 16.68650	& 18.39020 & 6.46258\\
		593--594* & 593.7716675    & 1 & 2.98 &7.40397 & 6.17976 & 5.88553\\
		\hline
		\hline
	\end{tabular}
	\caption{Candidates surviving both the known line veto and the single interferometer veto. The table lists the sub-band where the candidate is found (column 1), the estimated frequency $f_0$ quoted as the arithmetic mean of the minimum and the maximum frequencies ($f_{\rm min}$ and $f_{\rm max}$) in the optimal HMM path (column 2), the number of frequency bins ($\Delta f_{\rm drift}$) between $f_{\rm max}$ and $f_{\rm min}$ (column 3), the estimated $a_0$ (column 4), the original score $S_\cup$ yielded by searching the whole data set (column 5), and the scores from searching the first and second half of the data separately (column 6 and 7). The resolutions of $f_0$ and $a_0$ are $5.787037 \times 10^{-7}$\,Hz and 0.01805\,s, respectively. The candidates marked with an asterisk survive the manual veto in Section \ref{sec:half_veto} and require further follow up.}
	\label{tab:candidates_veto2}
\end{table*}

\subsubsection{$T_{\rm drift}$ veto}
\label{sec:longer_coherent_time_followup}
In general we can categorize any survivors of the $T_{\rm obs}/2$ veto into four groups with reference to the optimal paths detected in the original search. The groups are defined in Table \ref{tab:longer_Tdrift}. We expect $S$ to increase, as the block length $T_{\rm drift}$ increases, as long as $T_{\rm drift}$ remains shorter than the intrinsic spin-wandering time-scale. One could therefore imagine vetoing a candidate whose optimal Viterbi path does not wander significantly, if increasing $T_{\rm drift}$ up to the observed wandering time-scale does not increase $S$. However, based on our experience analysing injections (see Section \ref{sec:veto_safety}), we adopt a more conservative approach to reduce the false dismissal rate from this veto step. Specifically, we veto a candidate whose optimal Viterbi path does not wander significantly, if increasing $T_{\rm drift}$ up to the observed wandering time-scale yields $S<S_{\rm th}$ (i.e. $S$ drops below threshold) and the optimal paths returned for the two $T_{\rm drift}$ values do not match. For a candidate whose optimal Viterbi path does wander significantly, we do not expect $S$ to increase with $T_{\rm drift}$, if the intrinsic spin-wandering time-scale is effectively shorter than $T_{\rm drift}$ already. Indeed, it is reasonable for a strongly wandering signal to disappear when tracked with longer $T_{\rm drift}$. On the rare occasion when this does happen, the candidate is likely to be a noise artifact. Candidates surviving the $T_{\rm drift}$ veto need to be followed up with more sensitive search pipelines (e.g. Cross-Correlation \cite{Whelan2015}).

\begin{table*}[!tbh]
	\centering
	\setlength{\tabcolsep}{8pt}
	\renewcommand\arraystretch{1.1}
	\begin{tabular}{lll}
		\hline
		\hline
		&Higher $S$ with longer $T_{\rm drift}$& Lower $S$ with longer $T_{\rm drift}$\\
		\hline
		Low spin wandering & Follow up with more sensitive method & Veto\\
		High spin wandering & Unlikely to happen & Follow up with more sensitive method \\
		& & guided by observed Viterbi path\\
		\hline
		\hline
	\end{tabular}
	\caption{Subsequent actions to be taken for survivors of the vetoes in Section \ref{sec:line_veto}--\ref{sec:half_veto} according to the amount of spin wandering and $S$-versus-$T_{\rm drift}$ trend observed by the HMM. }
	\label{tab:longer_Tdrift}
\end{table*}

In this search, the two survivors asterisked in Table \ref{tab:candidates_veto2} do not display strong spin wandering; they drift within three and one $f_0$ bins (see Figure \ref{fig:path} in Appendix \ref{sec:candi_path}). Hence we expect $S$ to increase at approximately the same $f_0$, as $T_{\rm drift}$ increases all the way up to $T_{\rm obs}$. In fact we find that it suffices to consider $T_{\rm drift}=20$\,d. The original and follow-up results are recorded in Table \ref{tab:follow-up}. For $T_{\rm drift}=20$\,d, no path is detected with $S>S_{\rm th}$ at sub-bands 449\,Hz and 593\,Hz. The optimal Viterbi paths returned from $T_{\rm drift}=10$\,d and 20\,d are different in each of the two sub-bands, with an absolute discrepancy $\gtrsim 0.02$\,Hz and $\gtrsim 1.03$\,s for estimated $f_0$ and $a_0$, respectively. Normally the absolute uncertainties in the estimated values of $f_0$ and $a_0$ are less than 0.001\,Hz and 0.02\,s, respectively (see more details in Section \ref{sec:scox1-para} and Section IV B of Ref. \cite{Suvorova2016}). Hence we do not see any evidence of a real astrophysical signal in these two outliers.

\begin{table}
	\centering
	\setlength{\tabcolsep}{7pt}
	\renewcommand\arraystretch{1.1}
	\begin{tabular}{l  l l l }
		\hline
		\hline
		$T_{\rm drift}$ & Quantity & 449--450\,Hz & 593--594\,Hz\\
		\hline
		10\,d & $S$ & 7.38701  &  7.40397\\
		& $f_0$ (Hz)& 449.8116936 & 593.7716675\\
		& $a_0$ (s)& 2.73 &2.98\\
		\hline
		20\,d &  $S$ & 6.93366  & 6.93900\\
		& $f_0$ (Hz)& 449.7891863&  593.6174193\\
		& $a_0$ (s) & 1.70  & 1.79\\
		\hline
		\hline
	\end{tabular}
	\caption{Final-step follow-up with longer $T_{\rm drift}=20$\,d in two 1-Hz sub-bands containing the survivors from Section \ref{sec:half_veto}. The top and bottom halves of the table correspond to $T_{\rm drift}=10$\,d and 20\,d, respectively. The estimated $f_0$ is quoted as the arithmetic mean of min $f_0(t)$ and max $f_0(t)$ for the optimal Viterbi path. The follow-up score $S$ with $T_{\rm drift}=20$\,d is always below $S_{\rm th}=7.34$ and lower than the original score. The resolutions of $a_0$ and $f_0$ are 0.01805\,s and $5.787037 \times 10^{-7}$\,Hz, respectively for both $T_{\rm drift}=10$\,d and 20\,d.}
	\label{tab:follow-up}
\end{table}

\subsection{Veto safety}
\label{sec:veto_safety}
The four-step veto procedure is verified with four synthetic signals injected into 120\,d of Initial LIGO S5 data recolored to Advanced LIGO O1 noise and 200 signals injected into 130\,d of O1 data. The signals feature low spin wandering, drifting within one to four $f_0$ bins during the full observation. We do not inject signals into the sub-bands contaminated by known noise lines, so these 204 signals survive the first veto step in Section \ref{sec:line_veto} automatically. Only two out of the 204 injections are vetoed after the four steps described in Section \ref{sec:line_veto}--\ref{sec:longer_coherent_time_followup}, yielding a false dismissal rate $< 1\%$ and demonstrating that detectable spin-wandering signals are not commonly rejected. The two vetoed injections are rejected by the $T_{\rm obs}/2$ veto. They return a slightly higher $S$ value than $S_\cup$ (one in the first half, the other in the second), with $(S - S_\cup)/S_\cup \leq 3\%$ and $S_\cup\lesssim 10$ (i.e. $< 50\%$ higher than $S_{\rm th}$). In other words, the two false dismissals happen when both $(S - S_\cup)/S_\cup$ and $S_\cup$ are small. By contrast, three out of the four candidates vetoed in Table \ref{tab:candidates_veto2} (Section \ref{sec:half_veto}) return $(S - S_\cup)/S_\cup > 10\%$ (with $8< S_\cup<16$), and the other returns $S - S_\cup =0.35$ with $S_\cup>20$ (i.e.  $175\%$ higher than $S_{\rm th}$). Hence the four vetoed candidates in Table \ref{tab:candidates_veto2} fail the $T_{\rm obs}/2$ veto more strongly and are unlikely to be false dismissals.

Twelve examples of the synthetic signals surviving the vetoes described in Section \ref{sec:line_veto}--\ref{sec:longer_coherent_time_followup} are listed in Table \ref{tab:safety}. 

\begin{table*}
	\centering
	\setlength{\tabcolsep}{4pt}
	\renewcommand\arraystretch{1.1}
	\begin{tabular}{lllllllllll}
		\hline
		\hline
		Data & ${f_0}_{\rm inj}$ (Hz) & ${a_0}_{\rm inj}$ (s)& ${h_0}_{\rm inj}$ ($10^{-25}$)& ${\cos \iota}_{\rm inj}$ & $S_\cup$ & $S_{\rm H1}$  & $S_{\rm L1}$ & $S_{\rm 1st\, half}$ & $S_{\rm 2nd\,half}$ & $S_{20\,\rm d}$\\
		\hline
		S5 & 64.5774908 & 0.81 & 9.58 &$-0.5936$ & 9.12097 & $<S_{\rm th}$ & 7.42935* & $<S_{\rm th}$ &7.67254& 11.7985 \\
		S5 & 102.2907797 &  2.47 & 9.81 &	$-0.7988$ & 20.81940 & 16.17190 & 12.00540 &	20.63850 & 17.38740 & 25.81390 \\
		S5 & 202.8863982 & 2.34&	11.25 & $-0.9205$ & 15.80950 & 18.54850 & 17.46390 & 19.15680 & 18.9102 & 21.4415 \\
		S5 & 254.6697757 & 3.03&	14.55&	0.0375& 12.50180 & 	$<S_{\rm th}$ &	9.27111 &	10.8953 &7.74954 &15.0849 \\
		O1 & 97.2345635 & 2.15 & 4.50 & 0.71935 & 9.76216 & $<S_{\rm th}$ &7.53014*&7.29089& 8.91108&9.98727\\
		O1 &  132.1234568 &  0.70 & 4.80 & $-0.68154$  & 16.86500& 8.90286 & 8.63928 &13.29010 &13.30940 &19.54900\\
		O1 & 185.8094752 & 1.11 & 9.90 & 0.37952 & 19.05450 & 14.44080 & 12.70840& 18.07160 & 17.95120 & 20.34430\\
		O1 & 233.9125689 & 0.46 & 4.60 & 0.70917 & 16.71220 &  $<S_{\rm th}$ & 9.18889 & 12.25070  & 13.15180&18.02530\\
		O1 & 345.3456700 & 1.45 &7.00 & 0.71567& 14.09400& $<S_{\rm th}$& 9.15852 &10.10120&12.83390&14.72410\\
		O1 & 454.4563891 & 3.20 & 7.00 & $-0.86725$ & 9.03162 & 7.54074* &  $<S_{\rm th}$& $<S_{\rm th}$  & $<S_{\rm th}$  &9.06928\\
		O1 & 525.7096896 & 2.81 & 12.90 & 0.66578 & 11.55910 & 7.83362 &8.90156 & 11.35370&10.04660 &13.26430\\
		O1 & 635.6679700 & 1.98&10.00 & 0.72650 & 10.64010 & $<S_{\rm th}$&$<S_{\rm th}$&8.91769&9.13239&11.56240\\
		\hline
		\hline
	\end{tabular}
	\caption{Veto safety verification with synthetic signals. The table lists the data used for the injections (column 1), the injected signal parameters (column 2--5), the original score $S_\cup$ yielded by searching the whole data set with two interferometers and $T_{\rm obs}=10$\,d (column 6), the scores from searching H1 and L1 separately (column 7 and 8), the scores from searching the first and second half of the data separately (column 9 and 10), and the score with $T_{\rm obs}=20$\,d (column 11). A score is marked with an asterisk if it is above threshold, but the estimated frequency differs significantly from ${f_0}_{\rm inj}$ (i.e. wrong path returned). These twelve injections survive the four veto stages described in Section \ref{sec:line_veto}--\ref{sec:longer_coherent_time_followup}.}
	\label{tab:safety}
\end{table*}

\subsection{Strain upper limits}
\label{sec:upper_limit}
In the absence of a detection, we can place an upper limit on $h_0$ at a desired level of confidence (usually 95\%) as a function of $f_0$. 

A Bayesian analytic approach was adopted in the previous S5 sideband search for computing the strain upper limits \cite{ScoX1-S5}. However, the distribution of Viterbi path probabilities is hard to calculate analytically; Viterbi paths are correlated, and the nonlinear maximization step in the algorithm is hard to handle even within the context of extreme value theory (see Section III C in Ref. \cite{Suvorova2016}). Hence the Bayesian approach is hard to extend to the HMM sideband search. Instead, we adopt an empirical approach to set a frequentist upper limit as follows. We define $h_0^u$ such that the probability to detect a signal with $h_0 \geq h_0^u$ is greater than or equal to $u$, i.e. $\Pr(S\geq S_{\rm th}|h_0 \geq h_0^u) \geq u$. Hence with no detection we take the $h_0^{95\%}$ value plotted in Figure \ref{fig:sensitivity} (see Section \ref{sec:sensitivity}) as the frequentist 95\% confidence upper limit for electromagnetically restricted $\cos \iota$. It can be analytically converted to upper limits for unknown and circular polarizations using the scaling given by Equation (\ref{eqn:scaling_ul}).

\begin{figure*}
	\centering
	\scalebox{0.42}{\includegraphics{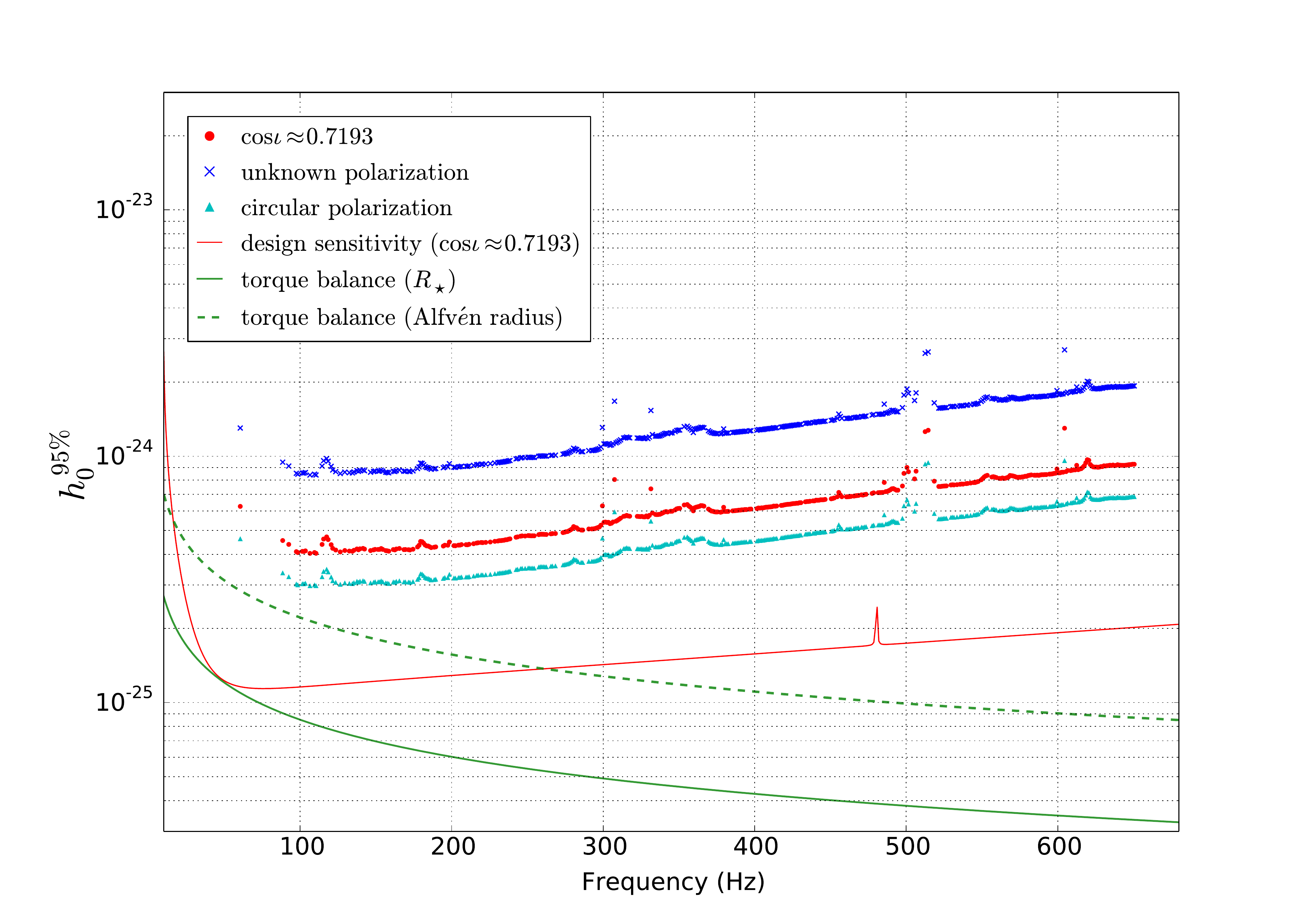}}
	\caption{Frequentist wave strain upper limits at 95\% confidence ($h_0^{95\%}$) as a function of signal frequency ($f_0$) assuming the electromagnetically restricted orientation $\iota = 44^\circ $ (red dots), unknown polarization with a flat prior on $\cos\iota$ (blue crosses), and circular polarization i.e. $\cos\iota=\pm1$ (cyan triangles). Each marker indicates the upper limit derived in the corresponding 1-Hz sub-band. Sub-bands with no marker are vetoed, e.g. contaminated by noise lines. The green solid and dashed curves indicate the theoretical torque-balance upper limits for LMXBs by taking $R_\star$ and the Alfv\'{e}n radius as the accretion-torque lever arm, respectively \cite{Bildsten1998}. The red curve indicates $h_0^{95\%}$ at the design sensitivity of Advanced LIGO \cite{aLIGO_design_sensi}, assuming $\iota = 44^\circ $ and $T_{\rm obs} = 2$\,yr.}
	\label{fig:UL}
\end{figure*}

Figure \ref{fig:UL} displays the upper limit derived from the O1 search combining data from H1 and L1 as a function of $f_0$. Each marker indicates $h_0^{95\%}$ in the corresponding 1-Hz sub-band. Bands that do not contain a marker are those containing a candidate vetoed in any of the four veto stages described in Section~\ref{sec:line_veto}--\ref{sec:longer_coherent_time_followup}. In total 180 out of 591 1-Hz sub-bands contain vetoed candidates (see Table \ref{tab:outlier_num}). The red dots correspond to assuming $\iota=44^\circ$, as inferred from radio observations \cite{Fomalont2001}. The blue crosses correspond to assuming unknown polarization and a flat prior on $\cos \iota$. The cyan triangles correspond to assuming circularly polarized signals (i.e. $\cos \iota=\pm1$). At 106\,Hz, the lowest 95\% confidence upper limits are $h_0^{95\%} = 4.0\times10^{-25}$, $8.3\times10^{-25}$ and $3.0\times10^{-25}$ for electromagnetically restricted $\cos \iota$, unknown polarization, and circular polarization, respectively. Hence the electromagnetically restricted prior and circular polarization assumptions improve upon the upper limits for unknown polarization by factors of 2.08 and 2.77, respectively.

As a further check, we compare the frequentist Viterbi upper limit to the frequentist $\mathcal{C}$-statistic upper limit. We run injections in six 1-Hz sub-bands in the best 10-day stretch of the real O1 interferometer data, starting from 110\,Hz, 257\,Hz, 355\,Hz, 454\,Hz, 550\,Hz and 649\,Hz, and search for them with the $\mathcal{C}$-statistic sideband pipeline \cite{Sammut2014,ScoX1-S5}. The best 10-day data stretch is selected from O1 as follows \cite{Wette2009a,Abadie2010}. A figure of merit, proportional to the signal-to-noise ratio (SNR), is defined by $\sum_{I,J}[S_h(f_I)]_J^{-1}$, where $[S_h(f_I)]_J$ is the strain noise power spectral density at discrete frequency bin $f_I$ in the $J^\text{th}$ SFT, and the summation is over all SFTs in each rolling 10-day stretch in O1. The 10-day data stretch with the highest value of this figure over the 60--650\,Hz band is selected. We compare the values of $h_0^{95\%}$ from the $\mathcal{C}$-statistic to the values plotted in Figure \ref{fig:UL}. The results show that the frequentist 95\% confidence upper limits from the $\mathcal{C}$-statistic are 1.46--1.74 times larger than those achieved from the search described in this paper.

\section{Torque-balance upper limit}
In LMXBs the gravitational wave strain inferred from the torque-balance scenario can be expressed as a function of the spin frequency of the neutron star $f_\star$ and the the X-ray flux $F_X$ according to \cite{Bildsten1998,Wagoner1984,Sammut2014}
\begin{eqnarray}
	\label{eqn:h_EQ}
	\nonumber h_0^{\rm eq}  = &&5.5\times 10^{-27}\left(\frac{F_X}{10^{-8}{\rm erg\,cm^{-2}\,s^{-1}}}\right)^{1/2}\left(\frac{R_\star}{10\,\rm km}\right)^{3/4}\\
	&& \times \left(\frac{1.4 M_\odot}{M_\star}\right)^{1/4} \left(\frac{300\,\rm Hz}{f_\star}\right)^{1/2},
\end{eqnarray}
where $R_\star$ is the stellar radius and $M_\star$ is the stellar mass.\footnote{We assume that the system emits gravitational radiation via the mass quadrupole channel. The analogous equation for current quadrupole radiation is given in Ref. \cite{Owen2010}.} We now ask how $h_0^{\rm eq}$ compares to the results of the analysis in Section \ref{sec:O1_search}.

Let us take the electromagnetically measured $F_X=4 \times 10^{-7}$\,erg\,cm$^{-2}$\,s$^{-1}$ \cite{Watts2008} for Sco X-1 and the fiducial values $R_\star=10$\,km and $M_\star=1.4M_\odot$. We plot $h_0^{\rm eq}$ as a function of $f_0 =2f_\star$ in Figure \ref{fig:UL} (green solid curve). Near 106\,Hz, where the best $h_0^{95\%}$ is reported, we obtain $h_0^{\rm eq}\approx 8.3 \times 10^{-26}$, which is $4.8$, $10.0$, and $3.6$ times lower than $h_0^{95\%}$ for electromagnetically restricted $\cos\iota$, unknown polarization, and circular polarization, respectively. The design sensitivity of Advanced LIGO is expected to improve further about two-fold relative to O1 \cite{Abbott2016-detector}. The anticipated $h_0^{95\%}$ at the design sensitivity of Advanced LIGO is plotted as a function of $f_0$ in Figure \ref{fig:UL} as the red curve, assuming an electromagnetically restricted orientation ($\iota = 44^\circ $) and $T_{\rm obs} = 2$\,yr. Near 50\,Hz, $h_0^{95\%}$ reaches $h_0^{\rm eq}$. 

The green solid curve in Figure \ref{fig:UL} is somewhat conservative \cite{ScoX1-S5}. If we consider the Alfv\'{e}n radius to be the accretion-torque lever arm, instead of $R_\star$ as assumed in (\ref{eqn:h_EQ}), then $h_0^{\rm eq}$ increases by a factor of a few. The Alfv\'{e}n radius is given by \cite{Bildsten1997}
\begin{eqnarray}
\label{eqn:alfven_r}
R_A &=& \left(\frac{B_\star^4R_\star^{12}}{2GM_\star \dot{M}^2}\right)^{1/7}\\
\nonumber  &=& 35 \left(\frac{B_\star}{10^9 {\rm G}}\right)^{4/7} 
\left(\frac{R_\star}{10\,\rm km}\right)^{12/7} \\
\label{eqn:alfven_r_2}
&& \times \left(\frac{1.4 M_\odot}{M_\star}\right)^{1/7} \left(\frac{10^{-8}M_\odot\,{\rm yr}^{-1}}{\dot{M}}\right)^{2/7} {\rm km},
\end{eqnarray}
where $B_\star$ is the magnetic field of the star, $G$ is Newton's gravitational constant, and $\dot{M}$ is the accretion rate. The neutron stars in LMXBs have $\dot{M}$ ranging from $\sim 10^{-11}M_\odot\,{\rm yr}^{-1}$ to the Eddington limit $2\times10^{-8}M_\odot\,{\rm yr}^{-1}$ \cite{Ritter2003,Sammut_thesis}, and weak magnetic fields in the range $10^8\,{\rm G} \lesssim B_\star \lesssim 10^9$\,G \cite{Bildsten1998,Patruno2012,Sammut_thesis}. To estimate the maximum magnitude of the effect, we substitute $\dot{M}=10^{-8}M_\odot\,{\rm yr}^{-1}$ and $B_\star=10^9$\,G in Equation (\ref{eqn:alfven_r_2}). The resulting $h_0^{\rm eq}$ is shown as the green dashed curve in Figure \ref{fig:UL}, giving $h_0^{95\%} \approx 2 h_0^{\rm eq}$ for electromagnetically restricted $\cos\iota$. At the design sensitivity of Advanced LIGO, we expect $h_0^{95\%} < h_0^{\rm eq}$ in the band $30\,{\rm Hz}\lesssim f_0 \lesssim 250\,{\rm Hz}$.

\section{Conclusion}

We perform an HMM sideband search for continuous gravitational waves from Sco X-1 in Advanced LIGO O1 data from 60\,Hz to 650\,Hz. The analysis is computationally efficient, requiring $\lesssim 3\times 10^3$ CPU-hr. We see no evidence of gravitational waves. Frequentist 95\% confidence upper limits of $h_0^{95\%} = 4.0\times10^{-25}$, $8.3\times10^{-25}$, and $3.0\times10^{-25}$ are derived at 106\,Hz for electromagnetically restricted $\cos \iota$, unknown polarization, and circular polarization, respectively. The upper limits are derived from Monte-Carlo simulations of spin-wandering signals. They are $4.8$, $10.0$, and $3.6$ times larger than the stellar radius torque-balance limit $h_0^{\rm eq}$, and approach $h_0^{\rm eq}$ more closely, if we treat the Alfv\'{e}n radius as the accretion-torque lever arm. An analysis of two years of Advanced LIGO data at design sensitivity with this search will be able to constrain the Alfv\'{e}n radius lever-arm scenario at frequencies below 300\,Hz. The best existing Bayesian 90\% confidence median strain upper limit from the radiometer O1 search is $h_0^{90\%}=6.7\times10^{-25}$ at 135\,Hz \cite{Radiometer_O1}. It converts to 95\% confidence median and maximum upper limits $h_0^{95\%}=7.8\times10^{-25}$ and $h_0^{95\%}=1.0\times10^{-24}$, respectively in the sub-band 134--135\,Hz \cite{MessengerNote}, which are comparable to the results for unknown polarization presented here.\footnote{The value of $h_0^{95\%}$ from the present search for unknown polarization is 6\% higher and 17\% lower than the median and maximum $h_0^{95\%}$ values from the radiometer search, respectively \cite{Radiometer_O1}. A direct comparison of the best quoted limits from the present search and the radiometer search is complicated by the different approaches of reporting upper limits. The present search returns the optimal Viterbi path (i.e. one upper limit) in each 1-Hz sub-band, while the radiometer search reports a range of upper limits.} Although these results are similar in sensitivity, this is the first analysis that searches over the projected semi-major axis of the binary orbit within the uncertainty of the electromagnetic measurement, while taking into account the effects of spin wandering over $T_{\rm obs}$. The spin frequency of Sco X-1 has not been determined conclusively, and could also lie below 60\,Hz. In the future, it is hoped that the number of instrumental lines at low frequencies will be reduced, enabling analysis below 60\,Hz, where $h_0^{\rm eq}$ is higher and hence easier to reach. At the design sensitivity of Advanced LIGO, it is anticipated that $h_0^{95\%} $ can be improved further by a factor of 2--3, reaching $h_0^{\rm eq}$ near 50\,Hz. In addition to Sco X-1, the search can be applied to other X-ray binaries including Cygnus X-3, the next brightest X-ray source after Sco X-1, and sources like XTE J1751-305 and 4U 1636-536, which show periodicities in the X-ray light curves and may indicate r-mode oscillations \cite{WhitePaper,Mahmoodifar2013,Haskell2015}. 

\section{Acknowledgements}

The authors gratefully acknowledge the support of the United States
National Science Foundation (NSF) for the construction and operation of the
LIGO Laboratory and Advanced LIGO as well as the Science and Technology Facilities Council (STFC) of the
United Kingdom, the Max-Planck-Society (MPS), and the State of
Niedersachsen/Germany for support of the construction of Advanced LIGO 
and construction and operation of the GEO600 detector. 
Additional support for Advanced LIGO was provided by the Australian Research Council.
The authors gratefully acknowledge the Italian Istituto Nazionale di Fisica Nucleare (INFN),  
the French Centre National de la Recherche Scientifique (CNRS) and
the Foundation for Fundamental Research on Matter supported by the Netherlands Organisation for Scientific Research, 
for the construction and operation of the Virgo detector
and the creation and support  of the EGO consortium. 
The authors also gratefully acknowledge research support from these agencies as well as by 
the Council of Scientific and Industrial Research of India, 
Department of Science and Technology, India,
Science \& Engineering Research Board (SERB), India,
Ministry of Human Resource Development, India,
the Spanish Ministerio de Econom\'ia y Competitividad,
the  Vicepresid\`encia i Conselleria d'Innovaci\'o, Recerca i Turisme and the Conselleria d'Educaci\'o i Universitat del Govern de les Illes Balears,
the National Science Centre of Poland,
the European Commission,
the Royal Society, 
the Scottish Funding Council, 
the Scottish Universities Physics Alliance, 
the Hungarian Scientific Research Fund (OTKA),
the Lyon Institute of Origins (LIO),
the National Research Foundation of Korea,
Industry Canada and the Province of Ontario through the Ministry of Economic Development and Innovation, 
the Natural Science and Engineering Research Council Canada,
Canadian Institute for Advanced Research,
the Brazilian Ministry of Science, Technology, and Innovation,
International Center for Theoretical Physics South American Institute for Fundamental Research (ICTP-SAIFR), 
Russian Foundation for Basic Research,
the Leverhulme Trust, 
the Research Corporation, 
Ministry of Science and Technology (MOST), Taiwan
and
the Kavli Foundation.
The authors gratefully acknowledge the support of the NSF, STFC, MPS, INFN, CNRS and the
State of Niedersachsen/Germany for provision of computational resources. This is LIGO document LIGO-P1700019.

\appendix
\section{Hidden Markov model}
\label{sec:hmm}
An HMM is a finite state automaton defined by a hidden (unobservable) state variable $q(t)$ transitioning between values from the set $\{q_1, \cdots, q_{N_Q}\}$ and an observable state variable $o(t)$ taking values from the set $\{o_1, \cdots, o_{N_O}\}$ at discrete times $\{t_0, \cdots, t_{N_T}\}$. The automaton jumps between hidden states from $t_n$ to $t_{n+1}$ with probability
\begin{equation}
	\label{eqn:prob_matrix}
	A_{q_j q_i} = \Pr [q(t_{n+1})=q_j|q(t_n)=q_i]
\end{equation}
and is observed in the state $o_j$ with emission probability
\begin{equation}
	\label{eqn:likelihood}
	L_{o_j q_i} = \Pr [o(t_n)=o_j|q(t_n)=q_i].
\end{equation}
For a Markov process, the probability that the hidden path $Q=\{q(t_0), \cdots, q(t_{N_T})\}$ gives rise to the observed sequence $O=\{o(t_0), \cdots, o(t_{N_T})\}$ is given by
\begin{equation}
	\label{eqn:prob}
	\begin{split}
		P(Q|O) = & L_{o(t_{N_T})q(t_{N_T})} A_{q(t_{N_T})q(t_{N_T-1})} \cdots L_{o(t_1)q(t_1)} \\ 
		& \times A_{q(t_1)q(t_0)} \Pi_{q(t_0)},
	\end{split}
\end{equation}
where 
\begin{equation}
	\Pi_{q_i} = \Pr [q(t_0)=q_i]
\end{equation}
is the prior.
The most probable path $Q^*(O)= \arg\max P(Q|O)$ maximizes $P(Q|O)$ and gives the best estimate of $q(t)$ over the total observation.

In this application, we map the discrete hidden states one-to-one to the frequency bins in the output of a frequency-domain estimator $G(f)$ (see Section \ref{sec:matched_filter}) computed over an interval of length $T_{\rm drift}$, with bin size $\Delta f_{\rm drift} = 1/(2 T_{\rm drift})$. We can always choose an intermediate time-scale $T_{\rm drift}$ in between the duration of one SFT, $T_{\rm SFT}=30$\,min, and the total observation time $T_{\rm obs}$ in order to satisfy
\begin{equation}
	\left|\int_t^{t+T_{\rm drift}}dt' \dot{f_0}(t')\right| < \Delta f_{\rm drift}
\end{equation}
for all $t$.\footnote{Frequency-domain, continuous-wave LIGO searches operate on SFTs rather than the time series of the detector output \cite{Riles2013}.} We assume that the spin wandering caused by accretion noise in Sco X-1 follows an unbiased Wiener process, in which $f_0(t)$ experiences a random walk and stays within $\Delta f_{\rm drift}$ for a duration less than a conservatively chosen $T_{\rm drift}=10$\,d, based on the assumption that the deviation of the accretion torque from its average value flips sign on the time-scale of observed fluctuations in the X-ray flux \cite{Ushomirsky2000,ScoX1-S5}.\footnote{For constant spin up or spin down, we are able to track a maximum rate $|\dot{f_0}| = \Delta f_{\rm drift}T_{\rm drift}^{-1}=7\times 10^{-13}$\,Hz\,s$^{-1}$. By way of comparison, without considering accretion noise, the secular spin-down (or spin-up) rate of LMXBs satisfies $|\dot{f_0}|\lesssim 10^{-14}$\,Hz\,s$^{-1}$ \cite{Patruno2012}.} Assuming continuous frequency wandering (i.e. no neutron star rotational glitches), equation (\ref{eqn:prob_matrix}) simplifies to the tridiagonal form
\begin{equation}
	\label{eqn:trans_matrix}
	A_{q_{i+1} q_i} = A_{q_i q_i} = A_{q_{i-1} q_i} = \frac{1}{3},
\end{equation}
with all other entries vanishing. The emission probability can be expressed in terms of $G(f)$ as
\begin{equation}
	\label{eqn:emi_prob_matrix}
	L_{o(t) q_i} \propto \exp[G({f_0}_i)],
\end{equation}
where $G({f_0}_i)$ is the log likelihood that the gravitational-wave signal frequency $f_0$ (e.g. twice the spin frequency of the star) lies in the frequency bin $[{f_0}_i, {f_0}_i +\Delta f_{\rm drift}]$ during the interval $[t, t+T_{\rm drift}]$. As we have no advance knowledge of $f_0$, we choose a uniform prior, viz.
\begin{equation}
	\Pi_{q_i} = N_Q^{-1}.
\end{equation}

\section{Viterbi algorithm}
\label{sec:viterbi}
The classic Viterbi algorithm \cite{Viterbi1967} provides a recursive, computationally efficient route to computing $Q^*(O)$, reducing the number of operations to $(N_T+1)N_Q \ln N_Q$ by binary maximization \cite{Quinn2001}. At every forward step $k$ ($1\leq k \leq N_T$) in the recursion, the algorithm eliminates all but $N_Q$ possible state sequences, and stores the $N_Q$ maximum probabilities ($1\leq i \leq N_Q$)
\begin{equation}
	\delta_{q_i}(t_k) = L_{o(t_k)q_i} \mathop{\max} \limits_{1 \leq j \leq N_Q} [A_{q_i q_j}\delta_{q_j}(t_{k-1})].
\end{equation}
It also stores the previous-step states of origin,
\begin{equation}
	\Phi_{q_i}(t_k) = \mathop{\arg \max} \limits_{1 \leq j \leq N_Q} [A_{q_i q_j}\delta_{q_j}(t_{k-1})],
\end{equation}
that maximize the probability at that step. The optimal Viterbi path is then reconstructed by backtracking according to
\begin{equation}
	q^*(t_k) = \Phi_{q^*(t_{k+1})}(t_{k+1})
\end{equation}
for $0 \leq k \leq N_T -1$. A detailed description of the algorithm can be found in Section II D of Ref. \cite{Suvorova2016}. 

\section{$T_{\rm obs}/2$ veto survivors: optimal Viterbi paths}
\label{sec:candi_path}

\begin{figure*}[!tbh]
	\centering
	\scalebox{0.3}{\includegraphics{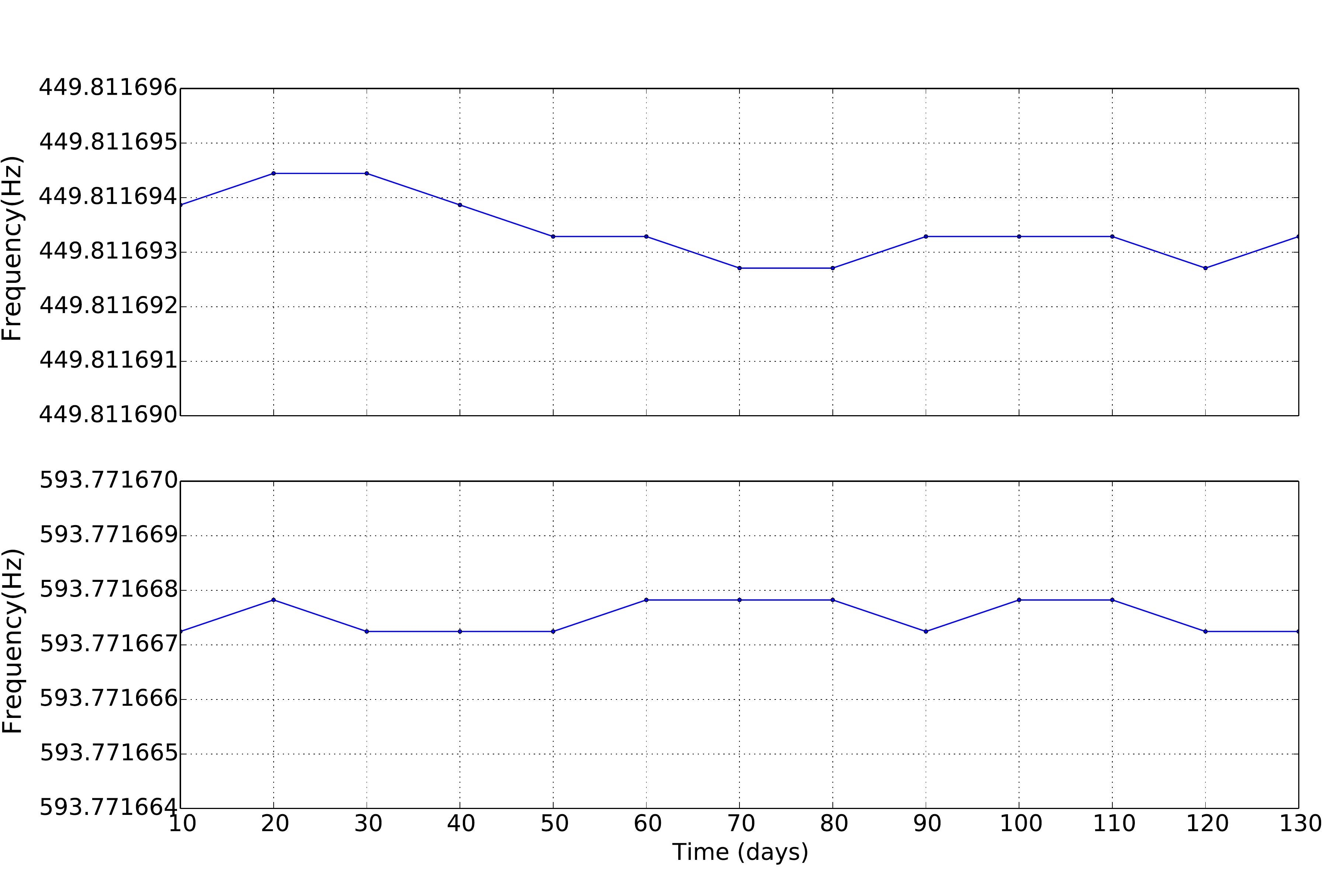}}
	\caption{Optimal Viterbi paths for the two survivors from Section \ref{sec:half_veto}.}
	\label{fig:path}
\end{figure*}

In the $T_{\rm drift}$ veto described in Section \ref{sec:longer_coherent_time_followup}, we categorize the two survivors according to their optimal paths detected in the original search. The optimal paths of the two survivors are plotted in Figure \ref{fig:path}, showing the estimated frequency $f_0$ as a function of time evaluated at the endpoint of each Viterbi step. The paths near 449\,Hz and 593\,Hz drift within three and one $f_0$ bins, respectively over $T_{\rm obs}$. They display low spin wandering.

\end{document}